\documentclass[a4paper]{JHEP3} 

\usepackage{amsmath}
\usepackage{amscd}
\usepackage{amsfonts}
\usepackage{amsthm}
\usepackage{bbm}

\def\nn         {\nonumber}

\def\calf         {{\cal F}}

\def\cali         {{\cal I}}

\def\caln         {{\cal N}}
\def\calo         {{\cal O}}
\def\calp         {{\cal P}}

\def\beal  {\begin{align}}
\def\bea  {\begin{eqnarray}}
\def\eea  {\end{eqnarray}}
\newcommand{\eq}[1]{\begin{equation}#1\end{equation}}
\newcommand\bbone{\ensuremath{\mathbbm{1}}}

\usepackage{dsfont}

\def\Re           {{\rm Re\hskip0.1em}}
\def\Im           {{\rm Im\hskip0.1em}}

\def\sqr#1#2{{\vcenter{\vbox{\hrule height.#2pt
 \hbox{\vrule width.#2pt height#1pt \kern#1pt \vrule width.#2pt}\hrule
 height.#2pt}}}}

\newcommand{\arXividhepth}[1]{\href{http://arxiv.org/abs/#1}{\tt arXiv:#1} [hep-th]}

\newcommand{\slashchar}[1]{\underline{#1}}

\title{Supersymmetric sources, integrability and generalized-structure compactifications}

\author{Paul Koerber  \\
        Max-Planck-Institut f\"{u}r Physik -- Theorie\\
        F\"{o}hringer Ring 6,  80805 M\"{u}nchen, Germany\\
        E-mail: \email{koerber@mppmu.mpg.de}}

\author{Dimitrios Tsimpis \\
Arnold Sommerfeld Center for Theoretical Physics\\
Department f\"{u}r Physik, Ludwig-Maximilians-Universit\"at\\
Theresienstr.\ 37, 80333 M\"unchen, Germany\\
E-mail: \email{tsimpis@theorie.physik.uni-muenchen.de}
}

\preprint{\arXivid{0706.1244}\\MPP-2007-66\\LMU-ASC 37/07
}

\abstract{In the context of supersymmetric compactifications of type II supergravity to
four dimensions, we show that orientifold sources can be compatible
with a generalized $SU(3)\times SU(3)$-structure that is neither
strictly $SU(3)$ nor static $SU(2)$. We illustrate this with explicit
examples, obtained by suitably T-dualizing known solutions on the six-torus.
In addition we prove the following integrability statements, valid under certain mild assumptions:
(a) for general type II supergravity backgrounds with orientifold and/or
D-brane generalized-calibrated sources, the source-corrected Einstein and dilaton equations of motion
follow automatically from the supersymmetry equations once the likewise source-corrected form equations
of motion and Bianchi identities are imposed; (b) in the special case of supersymmetric compactifications to
four-dimensional Minkowski space,
the equations of motion of all fields, including the NSNS three-form,
follow automatically once the supersymmetry and the Bianchi identities of the forms are imposed.
Both (a) and (b) are equally valid whether the sources are smeared or localized. As
a byproduct we obtain the calibration form for a space-filling NS5-brane.}


\begin{document}

\section{Introduction}

It has recently been appreciated that flux compactifications
 (for reviews see e.g.\ \cite{granareview,douglasreview,blumenhagenreview})
may be the right framework wherein to address long-standing issues which have
hitherto prevented string theory from making contact with realistic
low-energy physics. In this context, however, one faces the problem presented by
 the large number of possible string theory flux vacua. String theory is approximated at low-energies
by ten-dimensional supergravity,
possibly enriched by `stringy' objects such as orientifolds and D-branes,
and therefore a logical starting point would be to try to
explore the nature of supergravity solutions with fluxes turned on. To that end
it is useful to have as many generally-valid results as possible.

A natural language in which to recast the conditions for a supersymmetric $\caln=1$
flux compactification of type II supergravity to four-dimensional Minkowski space \cite{wittstruc,granaN1,jw}
is that of generalized complex geometry \cite{hitchin,gualtieri}.
More specifically: for a supersymmetric $\caln=1$ vacuum the six-dimensional
internal space must support two compatible generalized
complex structures,
one of which is integrable while the other one is not -- its nonintegrability being parameterized by the RR-fields.
The case of four-dimensional $\caln=1$
AdS vacua can also be described in the same formalism,\footnote{For a detailed analysis in the
language of ordinary $G$-structures, see \cite{behr,tsimpisads}. The last
reference contains the most general
form of supersymmetric compactifications to AdS$_4$ on manifolds
of $SU(3)$-structure, including a treatment of the equations of motion
and Bianchi identities.} although in this case neither of the generalized complex structures is
 integrable -- hinting at a role for `almost' generalized complex geometry \cite{granaN1,granascan}.

\subsection*{Equations of motion}

To obtain an $\caln=1$ vacuum the supersymmetry conditions
need to be supplemented by the form Bianchi identities
and equations of motion. General integrability results
ensure however that no further equations of motion need to be imposed.
Indeed, it was shown --
in \cite{tsimpisads} for IIA and
in \cite{gauntlettsusy} for IIB supergravity --
that the
Einstein equation and the dilaton equation of motion follow automatically from
the supersymmetry and the Bianchi identities and equations of motion of the
form-fields (for a discussion
in the context of M-theory see \cite{gauntlettsusyM}).

Upon adding D-brane and/or
orientifold sources to the supergravity Lagrangian
there will be a contribution to the Einstein and dilaton equation from the Dirac-Born-Infeld
terms, and to the Bianchi identities and equations of motion of the RR-fields from the Chern-Simons terms.
As we show in the present paper, provided the sources are supersymmetric
(or, equivalently, as we will explain in the following, generalized-calibrated), both contributions
exactly conspire so that supersymmetry together with the Bianchi
identities and equations of motion of the form-fields
still imply the Einstein and dilaton equations of motion\footnote{In the special case of the backgrounds of \cite{tsimpisads} with an O6 source,
it was verified in \cite{acharya} that the dilaton equation and the four-dimensional part
of the Einstein equation follow automatically,
provided that the sources are proportional to $\Re \Omega$ -- which in that context implies that
they are calibrated.}.

In hindsight it is obvious that the Einstein and dilaton equations of motion could not have followed
from the supersymmetry conditions alone, as this would have left no room for source contributions.
Nevertheless, it was noted in \cite{granascan} that, for the case of supersymmetric
compactifications to four-dimensional Minkowski vacua,
the equations of motion (but not the Bianchi identities) for the internal parts of the RR-fields do follow from the
supersymmetry equations. The rationale in this case is that the sources that would contribute to these equations
are forbidden because they would break the four-dimensional Poincar\'e symmetry. The proof is based on the integrability
of the generalized calibration conditions of the corresponding magnetic sources.
In the same way one would expect
the equation of motion for the NSNS $H$-field -- having as a source the fundamental string -- to follow from
the supersymmetry conditions. However, up to now it had not been possible to show this since the
equation of motion for the $H$-field does not fit well in the
generalized geometry framework \cite{tomasiello}. In the present paper
we will show that this equation does indeed follow from the supersymmetry equations and the Bianchi identities
-- much like the case of the RR fields. As a bonus, the proof also provides the calibration form
for the space-filling NS5-brane.

The situation can thus be schematically summarized as follows:

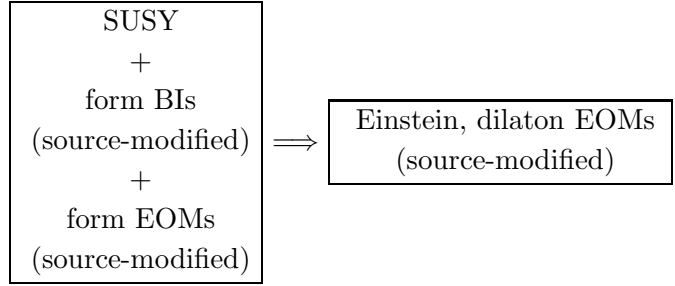
\begin{center}
\framebox{
{\begin{minipage}{3cm}\begin{center}
 SUSY\\ + \\ form BIs\\ (source-modified) \\ + \\ form EOMs\\ (source-modified)
\end{center}
\end{minipage}}}
$\Longrightarrow$ \framebox{
{\begin{minipage}{4.2cm}\begin{center}
Einstein, dilaton EOMs \\ (source-modified)
\end{center}
\end{minipage}}}
\end{center}

\noindent for general backgrounds, and

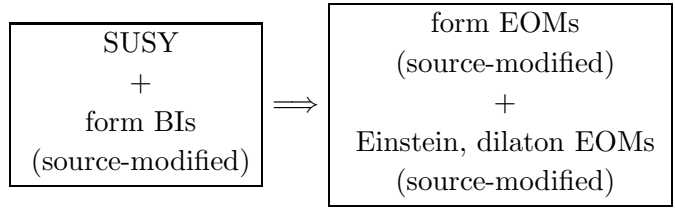
\begin{center}
\framebox{
{\begin{minipage}{3cm}\begin{center}
 SUSY\\ + \\ form BIs\\ (source-modified)
\end{center}
\end{minipage}}}
$\Longrightarrow$ \framebox{
{\begin{minipage}{4.2cm}\begin{center}
form EOMs \\ (source-modified) \\ + \\
Einstein, dilaton EOMs \\ (source-modified)
\end{center}
\end{minipage}}}
\end{center}

\noindent for compactifications to four-dimensional Minkowski space. Let us note that
we did not consider the possibility of adding a source term to the Bianchi identity of the $H$-field,
which would be generated by NS5-branes, although we expect our integrability results to be readily extendable
to include that case. We stress that
the proof remains equally valid whether the source terms are localized or not --
the main requirement being that the sources are generalized-calibrated.

The concept of a generalized-calibrated
D-brane was introduced in \cite{gencal,lucasmyth},
extending the work of \cite{gencalold}
to include a non-trivial gauge field $\calf$ on the world-volume.
A generalized-calibrated
D-brane extremizes its energy, rather than its volume
(which is the case for an ordinary calibration), and therefore corresponds to a (static)
solution of the equations of motion. Moreover, it was shown in \cite{gencal,lucasmyth}
that the calibration conditions
are equivalent to the requirement that the D-brane preserve the supersymmetry of the background.
The D-branes originally considered in \cite{gencal,lucasmyth} were localized;
the generalization of the calibration conditions
to the case of smeared D-brane/orientifold sources is straightforward.
The superpotential for the moduli space of generalized calibrations was introduced
in \cite{lucasuppot}, the deformation theory further studied in \cite{branedeform},
and networks of calibrated D-branes in \cite{jarahluca}.

\subsection*{Orientifolds}

Independent of whether the vacuum is supersymmetric or not, in the case of
flux compactifications of ten-dimensional supergravity to
Minkowski space there exists a no-go theorem \cite{nogo,gkp} which (under certain assumptions such
as the absence of higher-order derivative corrections) requires
the presence of sources with negative tension. In string theory,
such sources are indeed available: the orientifolds.

The supersymmetry conditions for orientifolds
in terms of the two generalized complex structures of the background, were
first studied in \cite{grimm}. In that reference,
the supersymmetry conditions were extrapolated from the analogous conditions in the (warped) Calabi-Yau case.
Subsequently, they were used in \cite{granascan} to argue that orientifolds would only be
compatible with either strict $SU(3)$-structure, or
static $SU(2)$-structure.
This would then imply that
the most interesting cases from the point-of-view of generalized
complex geometry -- i.e.\ solutions with dynamic $SU(3)\times SU(3)$-structure which is
neither strictly $SU(3)$ nor
static $SU(2)$, but interpolates between the two -- would effectively be excluded for flux compactifications.

In this paper we will derive and confirm the conditions of \cite{grimm}
from a world-sheet perspective,  as was done earlier for the Calabi-Yau
case in \cite{oriCY}. We will however also show that the
argument of \cite{granascan} is too restrictive, and that supersymmetric orientifolds {\em can} be
compatible with a dynamic $SU(3)\times SU(3)$-structure.
Moreover starting from a so-called type $B$ (or ``warped Calabi-Yau'') solution on a torus
and performing two T-dualities, we  will provide explicit
examples on nilmanifolds.
A non-geometric background with dynamic $SU(3)\times SU(3)$-structure and orientifold sources,
appeared recently in \cite{nongeommarch} (see also \cite{micu}).

This paper is organized as follows. After a brief review of generalized
geometry and supersymmetric vacua in the next section, we come to the orientifold
analysis in section \ref{orientifolds}. Our results concerning the integrability
of the supersymmetry equations in the presence of sources are contained in section \ref{integrability}.
The source-corrected equation of motion for the NSNS three-form as well as
the NS5 calibration form, are derived in section \ref{heom}.
We conclude in section \ref{conclusions}.
Many useful technical details can be found in the appendices.

After this paper was posted on the hep-th archive, we were informed of a forthcoming
publication \cite{orist} with potential overlap with the present work.

\section{Supersymmetry and generalized complex geometry}
\label{susygcg}

This section is a brief review of $\mathcal{N}=1$ supersymmetric backgrounds in the
language of generalized geometry, and is included here mainly to establish notation and
conventions. For more details the reader is referred to appendices \ref{appa}, \ref{sugra}.
For an introduction to generalized complex geometry we refer to e.g.\
section 3 of \cite{granascan} or, for a more complete treatment, to the original
work of \cite{gualtieri}.

\subsection{Setup}

We will consider type II supergravity and, for most of the paper (with the exception of section \ref{integrability}),
we will make the following compactification
ansatz for the metric
\begin{equation}
\label{metricansatz}
d s^2= e^{2A(y)} \eta_{\mu\nu} dx^{\mu}dx^{\nu} + g_{ij}(y) dy^i dy^j \, ,
\end{equation}
where $e^A$ is the warp factor, $\eta_{\mu\nu}$ the four-dimensional Minkowski metric, and $g_{ij}$
the metric of the
six-dimensional internal space. Apart from the metric, type II supergravity also contains the dilaton $\Phi$,
the NSNS three-form $H$ and the RR-fields $F_{(n)}$. We will use the democratic formalism of \cite{democratic}
\footnote{As in \cite{lucasmyth}, we make the following changes with respect to \cite{democratic}: in IIB we take $H \rightarrow -H$ and in IIA $C_{(n-1)} \rightarrow (-1)^{\frac{n-2}{2}} C_{(n-1)}$.}
with a doubling of the number of RR-fields, so that $n=0,2,4,6,8,10$ in type IIA and $n=1,3,5,7,9$ in type IIB.
The additional RR fields then satisfy \eqref{Fduality}. For further details on our conventions on type II supergravity see appendix \ref{sugra}.
For the RR-fields, the most general ansatz compatible with four-dimensional Poincar\'e
invariance reads:
\eq{
\label{Fansatz}
F = \hat{F} + \text{vol}_4 \wedge \tilde{F} \, ,
}
with $\text{vol}_4$ the (warped) four-dimensional volume form.
In addition, the most general ansatz for $\caln=1$ supersymmetry in four dimensions is
\begin{equation}
\label{N1spinoransatz}
\begin{split}
\epsilon^{(1)}(y) = & \zeta_+ \otimes
\eta^{(1)}_{+}(y)+\zeta_-\otimes \eta^{(1)}_{-}(y)\ , \\
\epsilon^{(2)}(y) = &\zeta_+\otimes
\eta^{(2)}_{\mp}(y)+\zeta_-\otimes \eta^{(2)}_{\pm}(y)\ ,
\end{split}
\end{equation}
where the upper/lower sign is for type IIA/IIB respectively,
and $\zeta_-$, $\eta_-^{(1,2)}$ are the complex conjugates of
$\zeta_+$, $\eta_+^{(1,2)}$. For our detailed spinor conventions we refer to section \ref{spinorconv}.
We define $|a|^2 = |\eta^{(1)}|^2$ and $|b|^2=|\eta^{(2)}|^2$. As shown in \cite{lucasmyth},
supersymmetric D-branes require $|a|^2=|b|^2$, which
will also be the case for supersymmetric orientifolds as we demonstrate in section \ref{orientifolds};
 we will assume this to be the case in the rest of the paper.

{}From bilinears of the internal spinors $\eta^{(1,2)}$ one
can construct the $SO(6,6)$ pure spinors $\Psi_\pm$ as follows
\begin{equation}
\label{purespinors}
\begin{split}
\slashchar{\Psi_+} & = \eta^{(1)}_+ \eta^{(2)\dagger}_+ \, ,\\
\slashchar{\Psi_-} & = \eta^{(1)}_+ \eta^{(2)\dagger}_- \, ,
\end{split}
\end{equation}
where the underline replaces the Dirac slash as in \eqref{slash}.
Indeed, we can identify polyforms, i.e.\ sums of forms of different dimensions, with spinor bilinears by contracting
the indices with gamma-matrices. On the other hand,
the polyforms are also isomorphic to spinors of $SO(6,6)$
(up to a choice of the volume form),
with the $SO(6,6)$ Clifford action defined as in \eqref{spinoraction}.
These spinors are pure, i.e.\ they have a null space of maximal dimension; for $SO(6,6)$
this is equal to six.
Indeed, in the spinor bilinear picture the null space consists of the three
 annihilators of $\eta^{(1)}_+$ acting on
the left and the three annihilators/creators of $\eta^{(2)}_+$ acting on the right.

In \cite{granaN1} it was shown that the supersymmetry
variations of the fermions \eqref{susyvar} vanish for
the above ansatz -- so that the compactification
preserves $\caln=1$ supersymmetry -- if and only if
\begin{subequations}\label{bulksusy}
\begin{align}
\label{bulksusy1}
d_H \left(e^{3A-\Phi} \Im \Psi_1 \right) & = \frac{e^{4A}}{16} \tilde{F} \, , \\
d_H \left(e^{A-\Phi} \Re \Psi_1 \right) & = 0 \, , \\
d_H \left(e^{2A-\Phi} \Psi_2 \right) & = 0 \, .
\end{align}
\end{subequations}
In the above we have
normalized the internal spinors such that $|a|^2+|b|^2=2|a|^2=e^A$. Moreover,
 we set $\Psi_1=\Psi_\mp$, $\Psi_2=\Psi_\pm$ for type IIA/IIB respectively.
The twisted exterior derivative $d_H$ is given by $d_H = d + H \wedge$.
It was noted in \cite{lucasmyth} that the above equations correspond to
space-filling, domain wall and string-like D-branes respectively, indicating
a close relation between the background supersymmetry and its supersymmetric probes.

In the language of $G$-structures the internal manifolds above have a structure group
contained in $SU(3)$ -- since they have at least one
nowhere-vanishing spinor: $\eta^{(1)}$. The appearance of a second
invariant spinor $\eta^{(2)}$ translates to the statement that
the internal manifolds have $SU(3) \times SU(3)$-structure.
This terminology may be somewhat
confusing from the point-of-view of ordinary $G$-structures,
since the second spinor may or may not be different from the first one.


In fact, in six dimensions the most general
relation between the two spinors is
\begin{equation}
\eta_+^{(2)} = c \eta_+^{(1)} + W^i \gamma_i \eta_-^{(1)} \, .
\label{spinorrel}
\end{equation}
%
If $\eta^{(1)}$ and $\eta^{(2)}$ are everywhere parallel, i.e.\ $W=0$ and $c \neq 0$,  we say that we
have a {\em strict $SU(3)$-structure}; if the two spinors are everywhere orthogonal,
i.e.\  $c=0,W\neq 0$,
we have a {\em static $SU(2)$-structure}.
The interpolating, generic, case is called {\em dynamic $SU(3) \times SU(3)$-structure}
or {\em local $SU(2)$-structure}. In the latter case,
it is possible to have $c \neq 0, W \neq 0$ at generic points and either $c=0$ or $W=0$ at special points --
which, as will shortly become clear, means that the {\em type} (the lowest
form-dimension appearing in the
corresponding polyform) of one of the pure spinors changes.

The existence of a static $SU(2)$-structure implies that the internal manifold has $SU(2)$-structure
in the ordinary sense. On the other hand,
the existence of a dynamic $SU(3) \times SU(3)$-structure
does not generally impose any further topological constraints beyond the existence of an
$SU(3)$-structure in the ordinary sense \cite{gualtieri}, except if $W \neq 0$ everywhere
which leads again to $SU(2)$-structure.

In the orientifold examples of section \ref{texamples} of this
paper, we will be considering structures on nilmanifolds (see appendix
\ref{nilmanifolds} for a brief review)
that are constant in the basis of left-invariant one-forms. In particular, this
implies that $c \neq 0, W \neq 0$ everywhere.
{}From the discussion of the previous paragraph, it then follows
that in these examples the internal manifolds have $SU(2)$-structure in the ordinary sense.
This is of course not a surprise, as nilmanifolds are parallelizable and have in fact
a trivial $G$-structure (i.e.\ their structure group is the identity).

It follows that in the nilmanifold case there exists a
complete basis of invariant spinors allowing for an extended supersymmetry ansatz.
This does not necessarily lead to vacua with extended supersymmetry in four dimensions,
since the differential equations \eqref{bulksusy} have to be satisfied in addition.
This observation  highlights a point which is a frequent
source of confusion, and is therefore  worth emphasizing:
{\em the generalized-structure of a
supergravity solution refers to the spinor ansatz,
not to the topological $G$-structure of the internal manifold}.

\subsection{(Local) $SU(2)$-structure}

For local and static $SU(2)$, it will be convenient
to express the pure spinors in terms of $SU(2)$-structure quantities.
Following \cite{granascan},
we introduce a unimodular internal spinor $\eta_+$ and set
\begin{subequations}
\label{defabV}
\begin{align}
\eta_+^{(1)} & = a \eta_+ \, , \\
\eta_+^{(2)} & = b ( k_\Vert \eta_+ + k_\perp V^i \gamma_i \eta_-) \, ,
\end{align}
\end{subequations}
with $2||V||^2=|k_\Vert|^2+|k_\perp|^2=1$ and $|a|^2=|b|^2$.
Comparing with \eqref{spinorrel} we have
\begin{equation}
c = \frac{b k_\Vert}{a} \, ,  \qquad W=\frac{b k_\perp V}{\bar{a}} \, .
\end{equation}
This description is redundant,
so that we can choose $k_\Vert$ and $k_\perp$ real and positive and absorb their phases in $b/a$ and $V$
respectively. We then
have $k_\Vert=|c|$ and $k_\perp=\sqrt{2}||W||$.
We can also rotate $\eta_+$ so that $b=\bar{a}$, and therefore
 only the phase of $b/a=e^{i\theta}$ has physical meaning.
Let us define
\begin{subequations}
\label{defomega}
\begin{align}
\omega^{(1)}_{ij} & = i \eta_+^\dagger \gamma_{ij} \eta_+ \, , \qquad & \tilde{\omega}_{ij} & = i \tilde{\eta}_+^{\dagger} \gamma_{ij} \tilde{\eta}_+ \, , \hspace{2cm} \\
\Omega^{(1)}_{ijk} & = i \eta_-^\dagger \gamma_{ijk} \eta_+ \, , \qquad & \tilde{\Omega}_{ijk} & = i \tilde{\eta}_-^{\dagger} \gamma_{ijk} \tilde{\eta}_+ \, ,
\end{align}
\end{subequations}
where $\tilde{\eta}_+ = V^i \gamma_i \eta_-$. The somewhat asymmetric notation highlights the fact that, except in the case of static $SU(2)$-structure, $\tilde{\omega}$ and $\tilde{\Omega}$
are different from the corresponding quantities built from $\eta^{(2)}_\pm$:
\eq{
\label{defomega2}
|b|^2 \omega^{(2)}_{ij} = i \eta_+^{(2)\dagger} \gamma_{ij} \eta_+^{(2)} \, , \qquad b^2 \Omega^{(2)}_{ijk} = i \eta_-^{(2)\dagger} \gamma_{ijk} \eta_+^{(2)} \, .
}
With these definitions it follows that
\begin{subequations}
\label{4domega}
\begin{align}
\omega^{(1)} & = \omega - 2i gV \wedge g\bar{V} \, , \\
\tilde{\omega} & = -\omega - 2i gV \wedge g\bar{V} \, ,
\end{align}
\end{subequations}
where $gV$ is the one-form dual to the vector $V$.
The two-form $\omega$ satisfies $\iota_{V} \omega = \iota_{{\bar{V}}} \omega=0$. Moreover
\begin{subequations}
\label{4dOmega}
\begin{align}
\Omega^{(1)} & = 2 g V \wedge \Omega_2 \, , \\
\tilde{\Omega} & = - 2 gV \wedge \bar{\Omega}_2 \, ,
\end{align}
\end{subequations}
where
\begin{equation}
\Omega_{2ij} = i \tilde{\eta}_+^{\dagger} \gamma_{ij} \eta_+ \, ,
\end{equation}
so that $\iota_{V} \Omega_2=\iota_{\bar{V}} \Omega_2=0$.
Another useful expression we will need later on is
\eq{
\label{cV}
\gamma_{i_1i_2} \eta_+ = -i \omega^{(1)}_{i_1i_2} \eta_+ - \frac{i}{2} \Omega^{(1)}_{i_1 i_2 j} \gamma^j \eta_- \, .
}

With the above definitions we can reexpress the pure spinors \eqref{purespinors} as follows
\begin{subequations}
\label{purespinorsexpl}
\begin{align}
\Psi_+ & = \frac{|a|^2}{8} e^{-i\theta} e^{2g V \wedge g \bar{V}} \left(k_\Vert e^{i\omega} +i k_\perp \Omega_2\right) \, , \\
\Psi_- & = \frac{|a|^2}{4} g V \wedge \left(k_\Vert i\Omega_2 - k_\perp e^{i \omega} \right) \, ,
\end{align}
\end{subequations}
with $e^{i \theta} = b/a$. These relations can be inverted using \eqref{inversepurespinors}.
We can now see that the type (the lowest form-dimension in a polyform)
of $(\Psi_+, \Psi_-)$ is in general $(0,1)$.
At special points where $k_\perp=0$ or $k_\Vert=0$, it jumps to $(0,3)$ or $(2,1)$ respectively.

\section{Orientifolds}
\label{orientifolds}

We now come to the study of supersymmetric orientifolds and, in particular, their
compatibility with the different types of generalized structures defined in the
previous section. After deriving the action of the orientifold involution on the pure spinors,
we will argue that the claim of \cite{granascan} can be relaxed, and
supersymmetric orientifolds {\em can} be  compatible
with dynamic $SU(3)\times SU(3)$-structure.


\subsection{The orientifold involution}

An orientifold action $\calo$ is a composition of a reflection on the world-sheet
(denoted by $\Omega_p$)
exchanging the left-movers with the right-movers,
and a target-space involution $\sigma$ ($\sigma^2=1$ on bosonic fields) acting
 on the internal manifold. A factor $(-1)^{F_L}$, where $F_L$ is the
fermion number of the left-movers, is sometimes needed to ensure $\calo^2=\bbone$
on all states including spinors. Whether it appears or not depends on the
number of $+1$-eigenvalues of $\sigma$, which also determines the dimensionality of
the orientifold plane. This is the fixed point set of the involution which,
in our case, fills the four-dimensional space-time.
In detail, the orientifold projection is given by
\begin{subequations}
\label{orientifoldproj}
\begin{align}
\text{IIB}: \qquad \calo & = \Omega_p \sigma \quad (O5/O9) \, , & \calo & = \Omega_p (-1)^{F_L}\sigma \quad (O3/O7)\, ,  \\
\text{IIA}: \qquad \calo & = \Omega_p \sigma \quad (O6) \, , &\calo & = \Omega_p (-1)^{F_L}\sigma \quad (O4/O8) \, .
\end{align}
\end{subequations}
In our conventions the O6 projection does not contain a $(-1)^{F_L}$ factor:
see appendix \ref{spinorconv} for more details.

For the dilaton $\Phi$, metric $g$ and NSNS three-form $H$ to be invariant
under the total orientifold projection $\calo$, they have to transform under the involution as
\begin{equation}
\sigma^* \Phi = \Phi \, , \qquad \sigma^* g = g \, , \qquad \sigma^*H = -H \, .
\end{equation}
For the RR fields we need
\begin{subequations}
\label{sigmaRR}
\begin{align}
\text{IIB}: \qquad \sigma^* F & = -\alpha(F) \quad (O5/O9) \, , & \sigma^* F & = \alpha(F) \quad (O3/O7) \, ,  \\
\text{IIA}: \qquad \sigma^* F & = \alpha(F) \quad (O6) \, , & \sigma^* F & = -\alpha(F) \quad (O4/O8)\, ,
\end{align}
\end{subequations}
where the action of $\alpha$ on forms is defined in \eqref{alphaform}.
The orientifold is supersymmetric if and only
if the orientifold operator leaves the total supersymmetry
generator $\epsilon^{(1)}_L + \epsilon^{(2)}_R$ invariant.
Since $\Omega_p$ exchanges left- and right-moving supersymmetry generators, we have
\begin{subequations}
\begin{align}
\qquad \sigma^* \epsilon^{(1)} & = \epsilon^{(2)} \, , & \sigma^* \epsilon^{(2)} & = \epsilon^{(1)} &  & (O5/O9,O6)  \, , \hspace{1cm} \\
\qquad \sigma^* \epsilon^{(1)} & = -\epsilon^{(2)} \, , & \sigma^* \epsilon^{(2)} & = \epsilon^{(1)} &  &  (O3/O7,O4/O8) \, , \hspace{1cm}
\end{align}
\end{subequations}
where $(-1)^{F_L}$ is responsible for the sign difference between the two
lines. Note that using \eqref{refl}, at the orientifold plane locus we find
(with suitable orientation conventions) exactly the same formula as eq.~\eqref{branesusy} for D-branes, provided we set
$\calf=0$.
Plugging in the ansatz \eqref{N1spinoransatz},
we immediately see that $\zeta$ is forced by the orientifold action
to be the same in both lines of \eqref{N1spinoransatz} forbidding an $\caln=2$ ansatz based on different $\zeta$s
in the two lines\footnote{An $\caln=2$ ansatz based on a doubling of the internal invariant spinors is still possible. See e.g.\ the example in \ref{example1}.}.
Furthermore, taking \eqref{gammadecomp} into account together
with the fact that $\sigma^*$ contains an even/odd number of internal $\gamma$-matrices
in IIB/IIA respectively,
we arrive at the following simple action on the internal supersymmetry generators
\begin{subequations}
\label{sigmaspin}
\begin{align}
\text{IIB}: & & \sigma^* \eta^{(1)}_{\pm} & = \eta^{(2)}_{\pm} \, , & \sigma^* \eta^{(2)}_{\pm} & = \eta^{(1)}_{\pm} & & (O5/O9) \, ,  \\
            & & \sigma^* \eta^{(1)}_{\pm} & = -\eta^{(2)}_{\pm} \, , & \sigma^* \eta^{(2)}_{\pm} & = \eta^{(1)}_{\pm} & & (O3/O7) \, , \\
\text{IIA}: & & \sigma^* \eta^{(1)}_{\pm} & = \eta^{(2)}_{\mp} \, , & \sigma^* \eta^{(2)}_{\pm} & =  \eta^{(1)}_{\mp} & & (O6) \, , \\
            & & \sigma^* \eta^{(1)}_{\pm} & = - \eta^{(2)}_{\mp} \, , & \sigma^* \eta^{(2)}_{\pm} & = - \eta^{(1)}_{\mp} & & (O4/O8) \, .
\end{align}
\end{subequations}
{}From $\sigma^2=1$ it follows that for supersymmetric orientifolds,
just as for supersymmetric D-branes, we should have $|a|=|b|$.
Furthermore, using \eqref{purespinorrealprop}, we see that
\begin{subequations}
\label{sigmapurespinors}
\begin{align}
\text{IIB}: & & \sigma^* \Psi_+ & = \alpha(\bar{\Psi}_+) \, , & \sigma^* \Psi_- & = - \alpha(\Psi_-) & & (O5/O9) \, ,  \\
            & & \sigma^* \Psi_+ & = -\alpha(\bar{\Psi}_+) \, , & \sigma^* \Psi_- & = \alpha(\Psi_-) & & (O3/O7) \, , \\
\text{IIA}: & & \sigma^* \Psi_+ & = \alpha(\Psi_+) \, , & \sigma^* \Psi_- & = \alpha(\bar{\Psi}_-) & & (O6) \, , \\
            & & \sigma^* \Psi_+ & = - \alpha(\Psi_+) \, , & \sigma^* \Psi_- & = - \alpha(\bar{\Psi}_-) & & (O4/O8) \, ,
\end{align}
\end{subequations}
which agrees with the conjectured transformations
of \cite{grimm,granascan}.

As a consistency check, one can verify that using \eqref{sigmaRR}
 and \eqref{sigmapurespinors}, the equations \eqref{bulksusy}
as well as the Calabi-Yau condition \eqref{CYcond}
transform covariantly. Finally, one can readily see that the generalized metric induced
by $(\sigma(\Psi_+),\sigma(\Psi_-))$ in the way explained around \eqref{genmetric}, is $(\sigma(g),\sigma(b))=(g,-b)$.

\subsection{Compatibility of dynamic $SU(3)\times SU(3)$-structure with orientifolds}

Let us start with relation \eqref{spinorrel}
and solve for $\eta^{(1)}$ to obtain:
\begin{equation}
\eta_+^{(1)} = \bar{c} \eta_+^{(2)} - W^i \gamma_i \eta_-^{(2)} \, .
\label{spinorrelrev}
\end{equation}
On the other hand, using \eqref{sigmaspin} we see that in
type IIB \eqref{spinorrel} transforms under $\sigma$ as
\begin{equation}
\pm \eta_+^{(1)} = \sigma(c) \eta_+^{(2)} + \sigma({W})^{i} \gamma_i \eta_-^{(2)} \, ,
\end{equation}
where the upper/lower sign is for $O5/O9$ and $O3/O7$ respectively.
Comparing with (\ref{spinorrelrev}) we find
\begin{subequations}
\label{IIBcond}
\begin{align}
\sigma(c) & = \pm \bar{c} \, , \\
\sigma(W) & = \mp W \, .
\end{align}
\end{subequations}
By considering this relation at the
orientifold fixed plane we find that
$c = \pm \bar{c}$. Moreover $W$ must be
perpendicular to the $O5$, along the $O7$ respectively.
For $O3$ and $O9$ we find that $W=0$ at the fixed plane.
We conclude that a supersymmetric O3-plane,
just as a D3-brane, can only exist at
points where the type is $(0,3)$. It follows that static $SU(2)$-structure is incompatible with
O3-planes.

Let us now look at IIA, for which \eqref{spinorrel} transforms under $\sigma$ as
\begin{equation}
\pm \eta_-^{(1)} = \sigma(c) \eta_-^{(2)} + \sigma(W)^{i} \gamma_i \eta_+^{(2)} \, .
\end{equation}
The upper/lower sign is for $O6$, $O4/O8$ respectively.
Comparing this with the complex conjugate of \eqref{spinorrelrev}
\begin{equation}
\eta_-^{(1)} = c \eta_-^{(2)} + \bar{W}^i \gamma_i \eta_+^{(2)} \, ,
\end{equation}
we find
\begin{subequations}
\label{IIAcond}
\begin{align}
\sigma(c) & = \pm c \, , \\
\sigma(W) & = \pm \bar{W} \, .
\end{align}
\end{subequations}
By considering these relations at the $O4/O8$ fixed plane,
we see that we must have $c=0$. It follows that the case of $O4/O8$ is
incompatible with strict $SU(3)$-structure.

{}From
\eqref{IIBcond} and \eqref{IIAcond} we see that
\begin{equation}
\sigma(k_\Vert)=k_\Vert \, , \qquad \sigma(k_\perp)=k_\perp \, .
\end{equation}
Considering \eqref{IIBcond} and \eqref{IIAcond} {\em on the orientifold plane itself}
we find that the phase $b/a=e^{i\theta}$, defined in \eqref{defabV}, is completely fixed for O3,O5,O7 and O9.
This phase is commonly used to classify strict $SU(3)$ compactifications: $b/a=\pm i$ is called type B and
$b/a=\pm 1$ is called type C, and so we see here that this carries over to some extent.
Moreover, on the O3- and O9-plane we can only have type $(0,3)$ for the pure spinors $(\Psi_1,\Psi_2)$ and on the O4- and O8-plane only type $(1,2)$.
We stress again that off the orientifold plane there are no such restrictions. If we specialize however to e.g.\ constant structures on nil- or
solvmanifolds, these properties do carry over to the whole internal manifold.
We list these properties in table \ref{sumori}.
\begin{table}
\begin{center}
\begin{tabular}{|c|c|c|c|}
\hline
O-plane & $b/a=e^{i\theta}$ & $V$ & allowed types \\ \hline
O3 & $\pm i$ & NA & (0,3) \\
O4 & NA & $\Re V\!\!\!\perp, \Im V\Vert$ & (1,2) \\
O5 & $\pm 1$ & $V\!\!\!\perp$ & all \\
O6 & free & $\Re V\Vert, \Im V\!\!\!\perp$ & all \\
O7 & $\pm i$ & $V\Vert$ &all \\
O8 & NA & $\Re V\!\!\!\perp, \Im V\Vert$ & (1,2) \\
O9 & $\pm 1$ & NA & (0,3) \\ \hline
\end{tabular}
\end{center}
\caption{Properties of the $SU(3)\times SU(3)$-structure for the different orientifold planes. The
phase $b/a=e^{i\theta}$ and the vector $V$ were defined in {\protect \eqref{defabV}}. $V\!\!\!\perp$
means $V$ is orthogonal to the orientifold plane, while $V \Vert$ means it is along the plane.\label{sumori}}
\end{table}

We can explicitly work out the orientifold action in terms of the forms defined in \eqref{defomega} in two ways.
We can either start from \eqref{sigmaspin} and definitions \eqref{defomega},
or we can use \eqref{sigmapurespinors}
and \eqref{purespinorsexpl} instead. In both cases we find
\begin{subequations}
\label{IIBsigma}
\begin{align}
\sigma^* \omega & = \left(k_\Vert^2 - k_\perp^2\right) \omega + 2 k_\Vert k_\perp \, \Re \Omega_2 \, , \\
\sigma^* \Omega_2 & = -k_\Vert^2 \Omega_2 + k_\perp^2 \bar{\Omega}_2 + 2 k_\Vert k_\perp \omega \, ,
\end{align}
\end{subequations}
for IIB, and
\begin{subequations}
\label{IIAsigma}
\begin{align}
\sigma^* \omega & = -\left(k_\Vert^2 - k_\perp^2\right) \omega - 2 k_\Vert k_\perp \, \Re  \Omega_2 \, , \\
\sigma^* \Omega_2 & = k_\Vert^2 \bar{\Omega}_2 - k_\perp^2 \Omega_2 - 2 k_\Vert k_\perp \omega \, ,
\end{align}
\end{subequations}
for IIA. We can now
see precisely where the caveat in the
proof of \cite{granascan} lies: {\em the requirement
that $\omega$ and $\Omega_2$ should not mix under the orientifold involution is too strong}.

Defining $k_\Vert=\cos \phi$
and $k_\perp=\sin \phi$ with $0 \le \phi \le \frac{\pi}{2}$ for IIB,
$k_\Vert=\cos \left(\phi+\frac{\pi}{2}\right)$
and $k_\perp=\sin \left(\phi+\frac{\pi}{2}\right)$ with $-\frac{\pi}{2} \le \phi \le 0$ for IIA,
we find
\begin{subequations}
\begin{align}
\sigma^* \omega & = \cos 2 \phi \, \omega + \sin 2 \phi \, \Re \Omega_2 \, \\
\sigma^* \Re \Omega_2 & = \sin 2 \phi \, \omega - \cos 2 \phi \, \Re \Omega_2 \, \\
\sigma^* \Im \Omega_2 & = - \Im \Omega_2 \, .
\end{align}
\end{subequations}
This is a rotation, over an angle $\pi$, in the $(\omega, \Re \Omega_2, \Im \Omega_2)$-space.

\subsection{Examples from T-duality}
\label{texamples}

We will now illustrate the point made in the previous subsection, i.e.\ that orientifolds can be
compatible with a dynamic $SU(3)\times SU(3)$-structure, by considering two
explicit examples  obtained by T-duality from known solutions. The first of the two
has $\mathcal{N}=2$ supersymmetry, and is therefore somewhat trivial; the
second one has  $\mathcal{N}=1$.

\subsubsection{Example 1}
\label{example1}

We start from a compactification on the torus $T^6$ with an $O3$-plane and
imaginary self-dual $G_3$ --
a so-called type B solution --
and apply two T-dualities, ending up with a nilmanifold. This was first considered
in \cite{kachru}. The way to obtain a dynamic $SU(3)\times SU(3)$-structure is to
choose one of the T-duality directions `misaligned'
with the complex structure. For ease of
comparison we will start from one of
the T-dual solutions discussed in \cite{granascan}.

For a type B or `warped Calabi-Yau' solution we have a strict $SU(3)$-structure with pure spinors
following from \eqref{purespinorsexpl} in the limit $k_\perp=0$
\eq{
\Psi_1 = \Psi_+ = \frac{|a|^2}{8} e^{-i\theta} e^{i \omega} \, , \qquad \Psi_2 = \Psi_-=\frac{|a|^2}{8} i \Omega \, ,
}
where for compatibility with $D3/O3$-branes
we must have $e^{-i \theta}=\pm i$. The supersymmetry conditions read
(see e.g.\ \cite{granaSU3,granareview})
\begin{subequations}
\label{typeBeq}
\begin{align}
& d \Omega + 3 dA \wedge \Omega = 0 \, , \qquad H \wedge \Omega = 0 \, ,\\
\label{susyomega}
& d \omega + 2 dA \wedge \omega = 0 \, , \qquad H \wedge \omega = 0 \, , \\
\label{susyF1}
& d \Phi=\hat{F}_1 = 0 \, , \\
\label{susyF5}
& 4 dA = \pm e^{\Phi} \star \hat{F}_5 \, , \\
& H = \mp e^{\Phi} \star \hat{F}_3  \, .
\end{align}
\end{subequations}
The last condition (together with $\hat{F}_3 \wedge \Omega=0$ which follows from the first condition) can be
rephrased
as the well-known
statement that $G_3=\hat{F}_3 + i e^{-\Phi} H$ is imaginary (anti-)self-dual: $\star G_3 = \pm i G_3$.

We take the following explicit solution for $e^{-i \theta}=i$
\begin{subequations}
\label{ex1}
\begin{align}
\omega & = e^1 \wedge e^4 - (\cos \alpha \, e^5 + \sin \alpha \, e^3) \wedge e^2 + e^6 \wedge (\cos \alpha \, e^3 - \sin \alpha \, e^5) \, , \\
\Omega & = -(e^1 - i e^4)\wedge (\cos \alpha \, e^5 + \sin \alpha \, e^3 +i e^2) \wedge (e^6 - i \cos \alpha \, e^3 + i \sin \alpha \, e^5) \, , \\
H & = e^{3A} (e^1 \wedge e^3 \wedge e^6 + e^1 \wedge e^2 \wedge e^5) \, , \\
e^\Phi \hat{F}_3 & = -e^{3A} (e^2 \wedge e^4 \wedge e^5 + e^3 \wedge e^4 \wedge e^6) \, , \\
e^ \Phi \hat{F}_5 & = -4 \star dA \, ,
\end{align}
\end{subequations}
which can be  obtained by rotating $\Omega$ and $\omega$ of the example
on p.\ 50 of \cite{granascan}
by an angle $\alpha$ in the $(e^5,e^3)$-plane,
keeping $H, \hat{F}_3$ fixed. The vielbeins $e^i=e^{-A}dx^i$ satisfy $de^i + dA \wedge e^i=0$.
One can easily verify that this still solves \eqref{typeBeq} for all values of $\alpha$.

We now perform a T-duality in the
directions $x^5$ and $x^6$.
The transformation of the vielbein under a T-duality in the direction $l$ is given in
\cite{hassan}
and reads
\eq{
(e_T^a)_{i} = (Q^{-1}_+)^j{}_{i} e^a{}_j \, ,
}
with
\eq{
\label{Qmatrix}
Q^{-1}_+ = \left( \begin{array}{cc} -g_{ll}^{-1} & -g_{ll}^{-1}(g+b)_{li} \\ 0 & \bbone \end{array}\right) \, .
}
We can take the gauge choice $b = x^1 dx^3 \wedge dx^6 + x^1 d x^2 \wedge d x^5$ so that:
\begin{subequations}
\begin{align}
e_T^5 & = - e^A (dx^5 - x^1 dx^2) \, \Longrightarrow d (e^{-A} e_T^5) = e^{2A} e^1 \wedge e^2 \, , \\
e_T^6 & = - e^A (dx^6 - x^1 dx^3) \, \Longrightarrow d (e^{-A} e_T^6) = e^{2A} e^1 \wedge e^3 \, ,
\end{align}
\end{subequations}
and all other vielbein components remain unchanged.

{}From \eqref{maurercartan} we read off that we end up with nilmanifold $(0,0,0,0,12,13)$,
which is $(n \; 4.6)$ of table 4 of \cite{granascan}.
The T-dual vielbein is $e_T^5 = e^A e^5_L$,
$e_T^6 = e^A e^6_L$ and $e^i = e^{-A} e^i_L$ for $i=1,\ldots,4$.
Furthermore, in the flat coordinates corresponding to the T-dual vielbeins, $\eta^{(1)}$
remains unchanged while
$\eta^{(2)}$ undergoes a reflection in the 5 and 6 direction:
\eq{
\eta^{(2)}_{T+} = \gamma_{\underline{5}} \gamma_{(6)} \gamma_{\underline{6}} \gamma_{(6)} \eta^{(2)}_+ = - e^{i \theta} \gamma_{\underline{56}} \eta^{(1)}_+ \, .
}
Using \eqref{sigmaspin} one can easily check that
this relation corresponds to the action of an O5 orientifold along $5,6$.
{}From \eqref{cV} we can read off
\begin{subequations}
\label{ex1T1}
\begin{align}
& c_T = \sin \alpha & \Longrightarrow & & k_{\Vert T} & =\sin \alpha \, , \quad e^{i\theta_T}=1 \, , \\
& W_T = - \frac{\cos \alpha}{2} (e^1 -i e^4) & \Longrightarrow & & k_{\perp T}& =\cos \alpha \, , \quad gV_T = -\frac{1}{2}(e^1 - i e^4) \, ,
\end{align}
\end{subequations}
where we assume that $0 \ge \alpha \ge \frac{\pi}{2}$. For $\alpha=0$ we find static $SU(2)$-structure while for $\alpha=\frac{\pi}{2}$
we have strict $SU(3)$. {}From \eqref{4domega} and \eqref{4dOmega} we find
\begin{subequations}
\label{ex1T2}
\begin{align}
\omega_T & = - (\cos \alpha \, e^5_T + \sin \alpha \, e^3) \wedge e^2 + e^6_T \wedge (\cos \alpha \, e^3 - \sin \alpha \, e^5_T) \, , \\
\Omega_{2 T} & = (\cos \alpha \, e^5_T + \sin \alpha \, e^3 +i e^2) \wedge (e^6_T - i \cos \alpha \, e^3 + i \sin \alpha \, e^5_T) \, .
\end{align}
\end{subequations}
Finally, after the two T-dualities the dilaton, NSNS three-form and RR fluxes take the form
\begin{subequations}
\begin{align}
& e^{\Phi_T} = e^\Phi e^{2A} \, , \\
& H_T = 0 \, , \\
& e^{\Phi_T} F_{(3)T} = e^{3A} \left( e^3 \wedge e^4 \wedge e_T^5 - e^2 \wedge e^4 \wedge e_T^6 \right)-4 \, e^{A} \star_4 dA \, .
\end{align}
\end{subequations}
One can verify that this solves the supersymmetry equations \eqref{bulksusy}.
Moreover we find for the source
\begin{align}
d F_3^T = e^{-\Phi} (2 + \tilde{\nabla}^2_- (e^{-4A})) dx^1 \wedge dx^2 \wedge dx^3 \wedge dx^4 \, ,
\end{align}
where $\nabla^2_-$ is the Laplacian constructed from the unwarped metric in $(x^1,x^2,x^3,x^4)$.
This has indeed the right sign for an orientifold, as will be argued below in \eqref{nogo},
and indeed corresponds
to the O5-plane obtained by applying two T-dualities to the original O3-plane.

The ``modulus'' $\alpha$ appears in the pure spinors, but not in the metric nor in any of the form-fields.
These non-metric moduli were discussed in some detail in \cite{granaN2}. It was shown that for $SU(3)$-structure
they are in the vector representation, and were excluded by hand.
These moduli are signals of extended supersymmetry, indicating that the $\caln=1$ description is not
appropriate anymore. Indeed, different
pure spinors (and thus also different pairs $\eta^{(1)},\eta^{(2)}$ of ordinary spinors)
are possible for the same background.
Since the supersymmetry equations \eqref{susyvar} are linear, we can take an arbitrary linear combination
with independent four-dimensional spinors $\zeta,\zeta',\ldots$ and thus make an extended supersymmetry ansatz. For $\caln=2$
\begin{equation}
\begin{split}
\epsilon^{(1)} = & \zeta_+ \otimes
\eta^{(1)}_{+}+ \zeta'_{+} \otimes
\eta'_+\!\!{}^{(1)}+\zeta_-\otimes \eta^{(1)}_{-} +\zeta_-\otimes \eta'_-\!\!{}^{(1)}\ , \\
\epsilon^{(2)} = & \zeta_+\otimes
\eta^{(2)}_{\mp}+ \zeta'_+\otimes
\eta'_{\mp}\!\!{}^{(2)}+\zeta_-\otimes \eta^{(2)}_{\pm}+\zeta'_-\otimes \eta'_{\pm}\!\!{}^{(2)}\ ,
\end{split}
\label{n2ans}
\end{equation}
with $\zeta_+,\zeta'_+$ (and complex conjugates $\zeta_-$ and $\zeta'_-$) generating the four-dimensional supersymmetry.

As was already noted in \cite{granascan}, the present example does indeed have  $\caln=2$ supersymmetry.
Explicitly, in the above ansatz we can take
\eq{
\eta_+^{(1)} = \eta_+ \, , \qquad
\eta'_+\!\!{}^{(1)} = -\gamma_{\underline{53}} \eta_+ \, , \qquad
\eta_+^{(2)} = i \gamma_{\underline{56}} \eta_+  \, , \qquad
\eta'_+\!\!{}^{(2)} = i \gamma_{\underline{63}} \eta_+ \, ,
}
where $\eta_+$ is the internal spinor generating the $\omega$ and $\Omega$ of \eqref{ex1}.
The pure spinors built from \eqref{ex1T1},\eqref{ex1T2}
 can be obtained by taking $\zeta=\cos \alpha/2 \, \zeta_T$
and $\zeta'=\sin \alpha/2 \, \zeta_T$, with $\zeta_T$ generating the corresponding $\caln=1$ spinor ansatz.
Hence the dynamic $SU(3)\times SU(3)$-structure is rather trivial in this example, since the same background
can be equally well described either
by a strict $SU(3)$- or by a static $SU(2)$-structure. Rather, one should
use the $\caln=2$ ansatz (\ref{n2ans})  instead.

We now turn to our next example, which
 only has $\mathcal{N}=1$ supersymmetry.

\subsubsection{Example 2}

Let us consider the type IIB example on p.\ 51 of \cite{granascan}, which corresponds to an
$\mathcal{N}=1$ solution on the flat six-torus
$T^6$. We make the following coordinate transformation:
\begin{align}
\left(
\begin{array}{c}
x^4 \\
x^5\\
x^6
\end{array}
\right) \longrightarrow\mathrm{M}
\left(
\begin{array}{c}
x^4 \\
x^5\\
x^6
\end{array}
\right) \, , \quad \text{with} \,\, \mathrm{M}= \left(
\begin{array}{ccc}
1& 0 & 1 \\
1 &1 &0 \\
-2 & 1&1
\end{array}
\right)
~.
\label{coordch}
\end{align}
After the above transformation, the metric is no longer diagonal along the
$x^4,x^5,x^6$ directions:
\begin{align}
g_{ij}=\frac{e^{-2A}}{16} \left(
\begin{array}{ccc}
11& -5 & 1 \\
-5 &11 &1 \\
1 & 1& 3
\end{array}
\right)
~,
\end{align}
and the vielbein in these directions is
\begin{align}
E^a{}_{i}=\frac{e^{-A}}{4} \left(
\begin{array}{ccc}
1& 1 & -1 \\
-1 &3 &1 \\
3 & -1& 1
\end{array}
\right)
~.
\label{vielbs}
\end{align}
In addition, let us define $e^a=e^{-A}\delta_i^a dx^i$, which should not be confused with the vielbein that
takes the above non-diagonal form.
Equation (\ref{coordch}) is merely a coordinate transformation, so we still have a solution of eqs.~\eqref{typeBeq}.
We have in mind to perform two T-dualities along the direction $x^1$ and the {\em transformed} direction $x^6$
so let us focus on the part of the $SU(3)$-structure containing $e^{16}$:
\begin{align}
\omega&=\frac{1}{4}e^{16} +\dots \, , \nn \\
\label{ex2struc}
\Omega&=2 \, e^{16}\wedge W  +\dots \, ,
\end{align}
with
\begin{align}
W= \frac{1}{8} \left( e^4-e^5+i(e^2+e^3) \right)~.
\end{align}
Moreover, after the transformation the
NSNS three-form $H$ takes on the simple form
\begin{align}
H=-dx^{124}-dx^{135}+dx^{236}~,
\end{align}
so that with a gauge choice we can set
\begin{align}
b=x^2 dx^{14}+x^3 dx^{15}+x^2 dx^{36}~.
\end{align}

The action of the two T-dualities on the spinors can be immediately read off
from eqs.~\eqref{cV}, \eqref{ex2struc} and is given by
$\eta\rightarrow\eta_T$, where $\eta_T^{(1)}=\eta^{(1)}$, and
\begin{align}
\eta_T^{(2)}=\frac{1}{\sqrt{3}} \eta^{(1)}_+ + \frac{4}{\sqrt{3}} W_{i} \gamma^i \eta^{(1)}_-  \, .
\end{align}
In the above it is understood that $\gamma^i$ is defined using the original vielbein.
Comparing with \eqref{spinorrel},
we see that we
end up with dynamic $SU(3)\times SU(3)$-structure. As before we also have an O5-plane, which
in this case is along the directions 1 and 6.
To actually determine the T-dual vielbein and nilmanifold we have to work a little harder.
Proceeding similarly to the previous example,
the action of T-duality
on the vielbein is encoded in the matrices $Q_{+ \; l=1}^{-1}$, ${Q}_{+ \; l=6}^{-1}$, defined
in \eqref{Qmatrix}. The T-dual vielbein, $e_T^a$, is then given by
\eq{
e_T^a=E^a{}_i( {Q}_{+\;l=1}^{-1}\cdot {Q}_{+\;l=6}^{-1})^i{}_j dx^j~.
}
It follows that
\eq{
\begin{split}
d (g_{11} E^1{}_b e_T^b) & = \partial_i b_{k1}  dx^i \wedge dx^k \, , \\
d (g_{66} E^6{}_b e_T^b) & = \partial_i b_{k6}  dx^i \wedge dx^k \, .
\end{split}
}
Putting $e^1_L=g_{11} E^1{}_b e_T^b,e^6_L=g_{66} E^6{}_b e_T^b$ and $e^i_L=dx^i$ otherwise, we find,
after some further relabelling and changing signs,
the nilmanifold $(n \; 4.4)$ of table 4 of \cite{granascan}.

Concluding, we arrive at the following recipe for constructing
examples by T-duality. We start from a constant (up to a warp factor)
type B solution on the torus. Then we perform two T-dualities along, say $5$ and $6$, where to end up with
a dynamic $SU(3)\times SU(3)$-structure, we only need to make sure that both $\omega_{56}$ and
$\Omega_{56i} dx^i$ are non-zero. This amounts to choosing the T-dual directions ``misaligned''
with the $SU(3)$-structure. Furthermore we should have $H_{56i} dx^i=0$ since otherwise
we would end up with a non-geometric T-dual. The resulting nilmanifold then only depends on $H$
since we find $f^5{}_{bc}=H_{5bc}$ and $f^6{}_{bc}=H_{6bc}$. With two T-dualities we then find
O5-backgrounds on nilmanifolds from 4.4 in table 4 of \cite{granascan} on. Furthermore, it turns out that
by performing three T-dualities in this way one can only get strict $SU(3)$-structure.

\section{Integrability in the presence of calibrated sources}
\label{integrability}

In this section we show that -- under certain conditions  --
in a bosonic supersymmetric background
the Einstein equation as well as the dilaton equation of motion follow
from setting the supersymmetry variations of the fermions to zero,
provided the equations of motion and Bianchi identities of all form-fields
 are also imposed and the sources are calibrated. The conditions mentioned are that
 there is a time/space split and that the sources are static,
without any world-volume electric fields. Moreover, the time/space components
of the Einstein equation $E_{0i}=0$ have to be imposed by hand.
In the absence of sources,
this was already shown  in \cite{tsimpisads} for IIA and \cite{gauntlettsusy}
for IIB, so here we will focus on the contribution of the sources.

The proof relies on the fact that { the sources are generalized calibrated}.
This follows naturally from the fact that the source
should preserve the supersymmetry of the background. As we will see, the
generalization to the case of smeared sources is straightforward.

For concreteness, let us
consider a single localized supersymmetric
D-brane source with world-volume $\Sigma$ and world-volume
gauge field $\calf=P_\Sigma[b]+F$, such that  $d\calf=P_\Sigma[H]$. The
case of a localized orientifold source can be obtained from the present analysis by
replacing $T_p \rightarrow T_{Op}=-2^{p-5} T_{p}$ and setting $\calf=0$.
Moreover, since the whole argument depends linearly on
the sources, it can be readily extended to arbitrary sums of D-branes and orientifolds.

The action for a localized D-brane source is given by
\begin{align}
S_{Dp} = - T_p \int_\Sigma e^{-\Phi} \sqrt{P_\Sigma[g]+\calf} + T_p \int_\Sigma \frac{C +\tilde{C}}{2} \wedge e^\calf \, ,
\label{sourceaction}
\end{align}
where $C$ are the gauge potentials defined above \eqref{Fduality} and $\tilde{C}$ are their magnetic duals.
At the level of the equations of motion the duality constraint \eqref{Fduality} will identify both, but at the level
of the action they should still be considered as different\footnote{We thank Toine Van Proeyen for discussions on this point. This
subtlety let to a mistake of a factor of $2$ in \eqref{eomB} in the previous version of this paper.}. The second term on the right hand side -- which contributes to the equations of
motion and Bianchi identities of the RR-fields -- is the easiest to analyse,
so let us consider that one first.
To proceed we define a current $j_{(\Sigma,\calf)}$ associated to
the D-brane $(\Sigma,\calf)$ such that for any polyform $\phi$
\eq{
\int_\Sigma \phi \wedge e^\calf = \int_Y \langle \phi , j_{(\Sigma,\calf)} \rangle \, .
}
This current, introduced in this form in \cite{branedeform},
can be thought of as a pure spinor whose annihilator space is the generalized
tangent bundle $T_{(\Sigma,\calf)}$ associated to $(\Sigma,\calf)$ \cite{gualtieri}.
So we can associate a pure spinor with a single source.
{}From $d\calf=P_\Sigma[H]$ it follows that
\eq{
d_H j_{(\Sigma,\calf)} = 0 \, ,
}
so $j_{(\Sigma,\calf)}$ defines a generalized cocycle
in $H$-twisted cohomology \cite{jarahluca}. In the democratic formalism,
the RR part of the  action reads
\eq{
S_{\text{RR}} = - \frac{1}{2 \kappa_{10}^2} \frac{1}{4} \sum_n (-1)^n \int_Y F_{(n)} \wedge \star F_{(n)} + T_p \int_Y \frac{C+\tilde{C}}{2} \wedge \alpha(j_{(\Sigma,\calf)}) \, ,
}
which immediately leads to the source-corrected equations of motion and Bianchi identities \eqref{eomB}.

The first term on the right-hand side of
(\ref{sourceaction}) is more complicated. In fact, without some relation
between the two terms of the D-brane action, we cannot expect it to give
an exactly matching contribution to the Einstein and dilaton equations.
This relation is of course provided by
the calibration condition, which is equivalent  \cite{gencal,lucasmyth} to the
requirement that
the D-brane source should be supersymmetric \cite{kappa}.
In the conventions of \cite{lucasmyth}:
\eq{
\Gamma_{Dp} \epsilon_2 = \epsilon_1 \, ,
\label{branesusy}
}
with
\eq{
\Gamma_{Dp}=\frac{1}{\sqrt{-\det
(P[g]+{\cal
F})}}\sum_{2l+s=p+1}\frac{\epsilon^{\alpha_1\ldots\alpha_{2l}\beta_{1}\ldots\beta_{s}}}{l!s!2^l}{\cal
F}_{\alpha_1\alpha_2}\cdots{\cal
F}_{\alpha_{2l-1}\alpha_{2l}}\Gamma_{\beta_1\ldots\beta_{s}} \, .
}
Moreover,
\eq{
\label{kappainv}
\left(\Gamma_{Dp}(\calf)\right)^{-1}
=(-1)^{\frac{(p+3)(p+2)}{2}} \Gamma_{Dp}(-\calf)=-\alpha(\Gamma_{Dp}) \, .
}

\subsection{Calibration}

To show that a supersymmetric D-brane source is necessarily
calibrated we proceed along the lines of \cite{gencal,lucasmyth}.
We extend that result to a more general setting and show
that when the D-brane is calibrated the Dirac-Born-Infeld action
reduces to an integration of the calibration form (appropriately twisted by $e^\calf$).

To make progress we must separate the time coordinate, so that the structure group
reduces as $SO(9,1) \rightarrow SO(9)$. Note that
this is a weaker condition than the
four-dimensional compactification ansatz  $SO(9,1) \rightarrow SO(3,1) \times SO(6)$
assumed in the other sections of the paper and in \cite{gencal,lucasmyth}.
The reason for making this time/space split is that there is a scalar representation in the
tensor decomposition of the $SO(9)$ spinor bilinear, while this is not the case for $SO(9,1)$
bilinears with spinors of the same chirality.
This allows us to define spinor norms, a prerequisite for the calibration argument which we will review in a moment.

In particular, the metric takes the form
\eq{
ds^2 = e^{2 A} dt^2 + \hat{g}_{ij} dx^i dx^j \, .
}
The supersymmetry parameters decompose as follows
\eq{
\epsilon_1 = \left(\begin{array}{c} 1 \\ 0 \end{array}\right) \otimes \hat{\epsilon}_1  \, , \qquad
\epsilon_2 = \left(\begin{array}{c} 1 \\ 0 \end{array}\right) \otimes \hat{\epsilon}_2  \quad \text{(IIB)} \, , \quad
\epsilon_2 = \left(\begin{array}{c} 0 \\ 1 \end{array}\right) \otimes \hat{\epsilon}_2 \quad \text{(IIA)} \, ,
\label{tencom}}
where $\hat{\epsilon}_{1,2}$ is the commuting $SO(9)$ part of the supersymmetry parameters.
The gamma-matrices decompose accordingly as
\eq{
\label{epsdecomp}
\Gamma_{\underline{i}} = \sigma_1 \otimes \hat{\gamma}_{\underline{i}} \, , \qquad \Gamma_{\underline{0}}= (i \sigma_2) \otimes \bbone \, , \qquad \Gamma_{(10)} =  \sigma_3 \otimes \bbone~,
}
with $\sigma_i$ the Pauli matrices, and $\hat{\gamma}_{\underline{i}}$ the 9-dimensional gamma-matrices.
We will take the latter to be real and symmetric. We
also need to impose a further condition, namely that $\Gamma_{Dp}$ splits as
\eq{
\Gamma_{Dp} = \Gamma_{\underline{0}} \, \Gamma_{Dp,\text{spatial}} = i \sigma_2 (\sigma_1)^p \otimes \hat{\gamma}_{Dp} \, ,
}
with $\hat{\gamma}_{Dp}$ purely spatial.
This will be the case if the D-brane configuration is static and there are no
electric world-volume fields. We note that this excludes some
interesting supersymmetric configurations such as the BIon \cite{BIon}.
It follows from \eqref{kappainv} that $\hat{\gamma}_{Dp}$ is symmetric, so that the norms of
the $SO(9)$ parts of the two
supersymmetry generators are equal:
$\hat{\epsilon}_1\!{}^T \hat{\epsilon}_1=\hat{\epsilon}_2\!{}^T \hat{\epsilon}_2=|a|^2$.

We are now ready to derive a calibration bound for the Dirac-Born-Infeld action
in a way completely analogous to \cite{gencal} (see \cite{thesis} for an earlier version).
We work purely in the spatial part and define
\eq{
\rho_{Dp} = \hat{\gamma}_{Dp} \sqrt{\det(P[\hat{g}]+\calf)} \, ,
}
so that
\eq{
\det(P[\hat{g}]+\calf) = \left(\rho_{Dp}\right)^T \rho_{Dp} \, .
}
We sandwich both sides between $\hat{\epsilon}_2\!{}^T$ and $\hat{\epsilon}_2$ and insert
a complete set $\bbone = \frac{1}{|a|^2} \sum_{\hat{\epsilon}'} \hat{\epsilon}' \, \hat{\epsilon}^{\prime T}$
to find
\eq{
\det(P[\hat{g}]+\calf) |a|^4 = \hat{\epsilon}_2\!{}^T \, \left(\rho_{Dp}\right)^T  \sum_{\hat{\epsilon}'}
\hat{\epsilon}' \, \hat{\epsilon}^{\prime T} \rho_{Dp} \hat{\epsilon}_2 \, .
}
Because $\hat{\epsilon}^{' T} \rho_{Dp} \hat{\epsilon}_2=\hat{\epsilon}_2\!{}^T  (\rho_{Dp})^T \hat{\epsilon}'$ the right-hand side
is in fact a sum of squares,
while in a  supersymmetric configuration only the term with
$\hat{\epsilon}_1\!{}^T \rho_{Dp} \hat{\epsilon}_2$ survives.
We have thus arrived at the advertised result that supersymmetric D-branes correspond
to generalized calibrated D-branes.
Indeed the Dirac-Born-Infeld part of the action, expanded around the supersymmetric configuration,
reduces as follows\footnote{If in addition one wishes to show that these D-branes minimize
the action, one would need to show that the remaining part of
the Dirac-Born-Infeld together with the Chern-Simons term is invariant under deformations. This amounts
to showing that $d_H \Psi=F$. Upon a $4+6$ split this indeed
follows from the background supersymmetry equations \eqref{bulksusy}.
For the minimal $1+9$ split we leave the analysis for future  work \cite{totalcal}.}
\eq{
\label{DBIred}
S_{\text{DBI},Dp}= -T_p \int_\Sigma P_\Sigma[\Psi] \wedge e^\calf + \calo(\text{cal}^2)
= -T_p \int_Y \langle \Psi, j_{(\Sigma,\calf)} \rangle + \calo(\text{cal}^2) \, ,
}
with
\eq{
\Psi = dt \wedge \sum_l \frac{e^{A-\Phi}}{l!|a|^2} \hat{\epsilon}_1\!{}^T \hat{\gamma}_{i_1 \ldots i_l} \hat{\epsilon}_2 \; dx^{i_1} \ldots dx^{i_l} \, .
\label{calform}
}
By $\calo(\text{cal}^2)$ we mean that the corrections to this calibrated configuration are quadratic in the
supersymmetry condition \eqref{branesusy}.

The contribution to the dilaton and Einstein equations of motion can now
be read off:
\begin{subequations}
\label{DBIcorrections}
\begin{align}
\frac{\delta S_{\text{DBI},Dp}}{\delta \Phi} & = T_p \, \langle \Psi, j_{(\Sigma,\calf)} \rangle \, , \\
\label{DBIcorrection2}
\frac{\delta S_{\text{DBI},Dp}}{\delta g^{N_1N_2}} & = - \frac{T_p}{2} \, \langle g_{N(N_1} dx^{N} \otimes \iota_{N_2)} \Psi,j_{(\Sigma,\calf)} \rangle \, .
\end{align}
\end{subequations}
In the above equations we use ten-dimensional notation, but the
reader should keep in mind that $\Psi$ transforms covariantly
only under time-independent coordinate transformations. Note in particular
that, as expected for static sources, the mixed time/space components on the
right-hand side of \eqref{DBIcorrection2} vanish.
The complete set of equations -- including the contribution of the sources -- for type II
supergravity is summarized in appendix \ref{sugra}.

So far we have assumed that
$j_{(\Sigma,\calf)}$  corresponds to a localized source, however the generalization
to smeared sources is immediate. We simply need to take \eqref{DBIred} as
the starting point for the Dirac-Born-Infeld action, in addition to imposing  the calibration condition.
The rest of the proof remains unchanged, whether
 the source is smeared or localized.

\subsection{Integrability}

Let us now come to the proof that supersymmetry
implies the dilaton and Einstein equations of motion,
provided that the form equations of motion and
Bianchi identities are satisfied.
In the absence of sources this has already been discussed
in detail in \cite{tsimpisads,gauntlettsusy}, so we only need to focus on the
contribution of the sources.
As already stressed, the crucial input for the proof to go through
is that the source terms are calibrated.

Let us use $j_{(n+1)}$ for the $(n+1)$-form part of $j_{(\Sigma,\calf)}$.
After some standard
(see for example \cite{tsimpisads})
gamma-matrix manipulations, taking the gravitino variation
(\ref{susyvar}) as well as
the identity
$8\nabla_{[M}\nabla_{N]}=R_{MNKL}\Gamma^{KL}$
 into account,
it follows that\footnote{We have found it
most convenient to perform the computation in the Einstein frame and then translate
back to the string frame.}
\begin{align}
E_{MN} \, \Gamma^N \epsilon-\kappa_{10}^2 e^{\Phi}
\frac{T_p}{2} \sum_n \slashchar{j_{(n+1)}}
\Gamma_{M}\calp_n\epsilon + \dots =0~,
\label{eintra}
\end{align}
where $E_{MN}=0$ is the Einstein equation without sources and the ellipsis
denotes terms that vanish under the projection onto the traceless symmetric part,
which we will apply in a moment.
To obtain the source term in the equation above, we also made use
of the following relation, which is a consequence of (\ref{eomB}),
\eq{
\slashchar{\nabla} Q_{M}=-\kappa_{10}^2
\frac{T_p}{4}\sum_n\slashchar{j_{(n+1)}} \Gamma_{M}\calp_n+\dots~,
}
where
\eq{
Q_{M}=
\frac{1}{16} \sum_n \slashchar{F_{(n)}} \Gamma_M \calp_n ~,
}
and the ellipsis denotes terms which do not depend on the sources.
To proceed, we can take the above equation for either $\epsilon_1$
or $\epsilon_2$, make the decomposition \eqref{epsdecomp} of the spinor and hit
on the left with $\hat{\epsilon}_1\!{}^T \Gamma_P$ or $\hat{\epsilon}_2\!{}^T \Gamma_P$ respectively.
Next, we project on the traceless and symmetric (in $M$ and $P$) part. For $M$ and $P$ purely spatial
we can take into account the following useful identity
\begin{align}
\hat{\epsilon}_1\!{}^T \hat{\gamma}_{(i} \, \slashchar{j_{(\Sigma,\calf)}} \, \hat{\gamma}_{j)} \hat{\epsilon}_2 =
(-1)^{n+1}\, 2\, e^{\Phi} |a|^2 \star \langle g_{k(i} dx^k \otimes \iota_{j)} \Psi, j_{(\Sigma,\calf)} \rangle \, ,
\end{align}
where both sides should be thought of as projected onto the traceless part. 
In this way, we arrive at
exactly the traceless part of the source-corrected Einstein equation (\ref{einstein}). Just as
in the absence of sources (see e.g.\ \cite{tsimpisads,gauntlettsusy}),
 the mixed time/space components of the
Einstein equation $E_{0i}=0$ have to be imposed by hand. Note that, as remarked below
\eqref{DBIcorrections}, the mixed time/space components of the source contribution
vanish identically for static sources.

The dilaton equation  can be treated similarly.
{}From the supersymmetry variations (\ref{susyvar}) it
 follows that
\begin{align}
\left(D-2\kappa_{10}^2 e^{\Phi}
{T_p}\sum_n(-1)^n\slashchar{j_{(n+1)}}\calp_n \right)
\epsilon=0~,
\label{hjk}
\end{align}
where $D=0$ is the dilaton equation in the absence of sources.
In the same way as above it correctly reproduces the trace of the source-corrected dilaton
equation (\ref{dilaton}).
Equation (\ref{hjk})  can be arrived at by noting that
\eq{
\nabla^{M}Q_{M}=-\kappa_{10}^2
\frac{T_p}{2}\sum_n(-1)^n\slashchar{j_{(n+1)}}\calp_n+\dots~,
}
where the ellipsis denotes source-independent terms.

Finally, the trace of the Einstein equation (\ref{einsteintrace})
follows from similar manipulations,
after using the dilaton equation to substitute for $\nabla^2\Phi$.

\section{The equation of motion for $H$}\label{heom}

For a compactification to four-dimensional Minkowski space, only space-time filling sources are allowed.
Indeed, sources that only partially fill the four-dimensional space-time would break Poincar\'e symmetry, while
instantonic sources are not
allowed in supergravity with Minkowskian signature. Such ``forbidden''
sources would appear
in the equations of motion of $\hat{F}$. As  shown in \cite{granascan} exactly these equations of motion
follow -- without source terms --
from the integrability of \eqref{bulksusy1}
\eq{
d_{-H} \star_6 \hat{F}=0 \, ,
}
where, as in the rest of this paper,
we assume that there are no NS5-brane sources so that $dH=0$.

The equation of motion for $H$ would have as a source the fundamental string which is similarly forbidden.
Since this equation does not fit very well in the language of generalized geometry, it is harder to show that
it also follows from supersymmetry. Nevertheless, a (tedious)
calculation shows that,  taking \eqref{eom3i} into account, supersymmetry implies:
\eq{
d \left[ e^{3A - 2 \Phi} \left( |a|^2 \omega^{(1)} - |b|^2 \omega^{(2)} \right)\right] = - e^{4A-2\Phi} \star_6 H
- 16 \, \left.\left(\alpha(\hat{F}) \wedge e^{3A-\Phi} \Im \Psi_1\right)\right|_{3}  \, ,
\label{hcalib}}
where $\omega^{(1)}$ and $\omega^{(2)}$ are constructed from $\eta^{(1)}$ and $\eta^{(2)}$ respectively as in
\eqref{defomega} and \eqref{defomega2}.
The calculation is quite similar to the one which shows that
\eqref{bulksusy1} follows from the supersymmetry
equations.

The term between the outer brackets on the left-hand side is
the calibration form for a space-filling NS5-brane.
The right-hand side then corresponds to its magnetic coupling to $H$ and its couplings,
via a Chern-Simons-like term,
to the RR-fields. This may be difficult to derive directly from the NS5-brane world-volume
action, which is rather complicated. It follows from the above that for strict $SU(3)$-structure and $|a|^2=|b|^2$,
a calibrated NS5-brane is not possible. However setting $a=0$ or $b=0$ (this
leads to the so-called
type A solutions, which only have NSNS-flux and
are common to type IIA, IIB and heterotic theory \cite{stromingerhet}),
one finds  space-filling supersymmetric NS5-branes
calibrating $\omega^{(1)}$ or $\omega^{(2)}$ respectively. These were studied in \cite{gauntNS5}.

Taking the exterior derivative of equation (\ref{hcalib}) leads to the
source-corrected equation of motion for
the $H$-field:
\eq{
d ( e^{4A - 2 \Phi} \star_6 H ) - e^{4A} \sum_n \star_6 \hat{F}_{(n+2)} \wedge \hat{F}_{(n)} +  16 \, (2 \kappa_{10}^2) \left. \left(e^{3A-\Phi} \Im \Psi_1 \wedge \alpha(j_{\text{total}})\right)\right|_4= 0 \, .
}
The contribution from the sources vanishes in some
common cases, like O3- and O5-planes, which may be the reason why it was not noted before.

We conclude that the equation of motion for $H$ is implied by the supersymmetry and
the Bianchi identities. The proof highlights the fact that the close relationship
between background supersymmetry conditions and calibrations also holds for NS5-branes.

\section{Conclusions}\label{conclusions}

We have seen that not only D-brane sources, but also supersymmetric orientifolds can be compatible
with a dynamic $SU(3)\times SU(3)$-structure. This opens up the possibility
for the construction of compactification manifolds which are
highly non-trivial from the generalized-geometry point-of-view.
Taking the integrability results of this paper into account we can summarize
the minimal conditions that a supersymmetric vacuum has to obey as in appendix \ref{allconditions}.
Unfortunately an extensive, although not exhaustive, scan of the nilmanifolds and solvmanifolds
has produced no explicit examples -- except for the ones that are T-dual to the six-torus.

As in \cite{granascan} we restricted
to left-invariant structures, i.e.~structures that are constant in terms of the left-invariant
one-forms, and orientifolds whose action takes a simple diagonal form in this basis.
Our results may be pointing to the fact
that these genuine generalized-structure backgrounds are very rare.
Alternatively, it may be
that nilmanifolds and solvmanifolds are simply not the right class to look for examples,
or the two simplifying assumptions we made within this class are too restrictive.
We should keep in mind that
the authors of \cite{granascan} only found a few examples of static $SU(2)$- and strict $SU(3)$-structure,
all of them on just two of these manifolds.

Although the equation of motion for the NSNS three-form does not fit very well in the
generalized-geometry framework, we were able to show that -- for compactifications
to four-dimensional Minkowski space -- it simply follows from the supersymmetry conditions and the
Bianchi identities. Not having to impose this equation as an extra condition,
should facilitate mathematical considerations concerning general properties of generalized vacua,
such as have recently appeared in \cite{tomasiello}. Furthermore we have established that the close connection
between the supersymmetry conditions of the background and the generalized calibrations
of supersymmetric probes, extends to the case of the NS5-brane.

Our integrability results show that the usefulness of generalized calibrations
extends beyond the probe approximation to fully back-reacting sources. Indeed,
having precisely these calibrated sources ensures that the source-corrected Einstein and dilaton equations still
follow from the supersymmetry conditions and the equations for the form fields.
Since the supersymmetry equations are much easier to analyse than the
equations of motion, our integrability results open up a host of new possibilities
for supergravity solutions with (smeared) sources. The potential phenomenological importance
of such vacua was recently noted in \cite{acharya}.

Of potential phenomenological importance is also the application of our results
 to $\mathrm{AdS}_5/\mathrm{CFT}_4$: five-dimensional AdS space can be thought of (in appropriate coordinates) as
four-dimensional warped Minkowski space, and therefore the relevant
strong integrability statement of the present paper applies. In many
physically interesting setups one would like to consider the addition of back-reacting sources
to the background, leading to source-modified Bianchi identities.
In the past several authors have checked on a case-by-case basis
\cite{nuneza, paredes, nunezb, nunezc} that once a supersymmetric $\mathrm{AdS}_5$ background with
supersymmetric sources satisfies the
source-modified Bianchi identities, the source-modified dilaton and Einstein equations follow.
Thanks to the results of the present paper, we now know that this is in fact a general result.

The study of four-dimensional AdS vacua from the point-of-view of generalized structure,
would also be an interesting avenue for future research.
Finally, it would be interesting to obtain an alternative
derivation of  the calibration form for the space-filling NS5-brane presented here,
directly from a world-volume analysis. Exploiting the connection between bulk supersymmetry
and calibrated probes, may lead to a better handle on
the complicated world-volume action for the NS5-brane.

\bigskip

\acknowledgments

P.K. wishes to thank Jerome Gauntlett and Toine Van Proeyen for discussions. We are also grateful to
Luca Martucci for useful remarks and proofreading,
and to Fernando Marchesano for making his latest paper available to us before publication.

\appendix

\section{Notation, conventions and useful formulae}
\label{appa}

In this appendix we explain in more
detail our conventions and notation, and we summarize
several useful technical points referred to in the main text.

\subsection{General}

Let $\alpha$ be the operator that reverses all the indices of a (poly)form
\eq{
\alpha(\phi)_{M_1\ldots M_n}=\phi_{M_n\ldots M_1} \, .
\label{alphaform}
}
The Mukai pairing between polyforms is defined as
\eq{
\langle \phi_1, \phi_2 \rangle = \phi_1 \wedge \alpha(\phi_2)|_{\text{top}} \, ,
}
where we select the top form.
The $\varepsilon$ tensor is given by $\varepsilon^{0 \ldots (D-1)}=-\varepsilon_{0 \ldots (D-1)}=1$. We define
the Hodge dual of a form as follows
\eq{
(\star \phi)^{M_1 \ldots M_l} = \frac{1}{\sqrt{|\det g|} \, (D-l)!} \varepsilon^{M_1 \ldots M_l N_1 \ldots N_{D-l}} \phi_{N_1 \ldots N_{D-l}} \, ,
}
and the contraction of a top form (or the top form part of a polyform) with the $\varepsilon$-tensor
\eq{
\phi|_\varepsilon = \frac{1}{D!} \phi_{M_1 \ldots M_{D}} \varepsilon^{M_1 \ldots M_{D}} = \sqrt{|\det g|} \; \star \phi \, .
}
We introduce the following notation for
the contraction of a (poly)form with gamma matrices:
\eq{
\label{slash}
\slashchar{\phi} = \sum_l \frac{1}{l!} \phi_{M_1 \ldots M_l} \Gamma^{M_1 \ldots M_l} \, .
}
For any form $A$ we have $\Gamma_{(10)} \slashchar{A}=\slashchar{\star\alpha(A)}$ and in particular for the RR fields, using \eqref{Fduality},
$\Gamma_{(10)} \slashchar{F}=\slashchar{F}$.

Throughout the text we will use the above definitions for both the total ten-dimensional space-time $Y$,
as, mutatis mutandis, for the internal manifold six-dimensional $M$.

\subsection{Spinors, ordinary \& generalized}
\label{spinorconv}

\subsubsection*{Ordinary spinors}

With the compactification ansatz \eqref{metricansatz}, the
ten-dimensional $\Gamma$-matrices decompose accordingly as
\begin{equation}
\Gamma_{\mu} = \tilde{\gamma}_{\mu} \otimes \bbone \, , \qquad \Gamma_i = \tilde{\gamma}_{(4)} \otimes \gamma_i \, ,
\label{gammadecomp}
\end{equation}
with $\tilde{\gamma}_{\mu}$ four-dimensional and $\gamma_i$ six-dimensional gamma-matrices, and
\begin{equation}
\tilde{\gamma}_{(4)} = i \tilde{\gamma}^{\underline{0123}} \, , \qquad
\gamma_{(6)} = - i \gamma^{\underline{123456}} \, ,
\end{equation}
the corresponding four-dimensional and six-dimensional chirality operators. The ten-dimensional
chirality operator reads
\begin{equation}
\Gamma_{(10)} = \tilde{\gamma}_{(4)} \otimes \gamma_{(6)} \, .
\end{equation}
We impose the following Majorana condition in ten dimensions
\begin{equation}
\label{Mcondition}
\epsilon = B_{(10)} \epsilon^* \, ,
\end{equation}
with $B_{(10)}= B_{(4)} \otimes B_{(6)}$, where $B_{(4)}$ and $B_{(6)}$ are used to impose the Majorana conditions in four
and ten dimensions
\begin{equation}
\label{Mcondition6d}
\zeta_{\pm} = B_{(4)} \zeta_{\mp}^* \, , \qquad\eta_{\pm} = B_{(6)} \eta_{\mp}^*  \, ,
\end{equation}
and  satisfy the defining relations
\begin{subequations}
\label{majorana4d6d}
\begin{align}
B_{(4)}^{-1} \tilde{\gamma}_{\mu} B_{(4)} & = \tilde{\gamma}_{\mu}^* \, , \\
B_{(6)}^{-1} \gamma_{i} B_{(6)} & = -\gamma_{i}^* \, .
\end{align}
\end{subequations}
It also follows,
as required for consistency, that $B_{(4)} B_{(4)}^*=B_{(6)} B_{(6)}^* = \bbone$.
Note that this consistency condition
does not allow for other choices of signs in \eqref{majorana4d6d}. {}From \eqref{gammadecomp}
we find
\begin{equation}
B_{(10)}^{-1} \Gamma_M B_{(10)} = \Gamma_M^* \, ,
\end{equation}
and, as again required for consistency,
$B_{(10)} B_{(10)}^*=\bbone$.
Note that if we define the usual charge conjugation matrix $C_{(10)}$ by
\eq{
C_{(10)} \Gamma_M C_{(10)}^{-1} = - (\Gamma_M)^T \, ,
}
and setting in addition $\bar{\epsilon}=\epsilon^\dagger \Gamma^{\underline{0}}$, we see that \eqref{Mcondition} can be
cast in more standard form $\bar{\epsilon} = \epsilon^T C_{(10)}$.
In ten dimensions there is another choice
for the matrix imposing the Majorana condition,
namely $\tilde{B}_{(10)} = \Gamma_{(10)} B_{(10)}$. This would lead to the introduction of
the operator $(-1)^{F_L}$ for the $O6$, as seems to be the usual convention. We will not make this choice
here as it would not be compatible with our
spinor ansatz \eqref{N1spinoransatz} for type IIA without some inconvenient sign changes.

A reflection in the $i$th internal direction is generated on spinors by
\begin{equation}
\label{refl}
\Gamma_{i} \Gamma_{(10)} = \bbone \otimes \gamma_i \gamma_{(6)} \, ,
\end{equation}
and with the above reality condition \eqref{Mcondition}
it converts Majorana spinors into Majorana spinors.
We note that $\Gamma_i \Gamma_{(10)} \Gamma_j \Gamma_{(10)}=-\Gamma_j \Gamma_{(10)} \Gamma_i \Gamma_{(10)}$
and $(\Gamma_i \Gamma_{(10)})^2 = -\bbone$, so that if $\sigma$ contains $l=9-p$ internal reflections
we have on spinors
\begin{equation}
\sigma^2 = (-1)^l (-1)^{\frac{l(l-1)}{2}} \bbone \, .
\end{equation}
If $\sigma^2=-\bbone$ we need to compensate in the orientifold projection with a factor of $(-1)^{F_L}$,
resulting in the choices of \eqref{orientifoldproj}.

\subsubsection*{Generalized spinors}

A generalized vector $\mathbb{X}=(X,a) \in T_M \oplus T_M^\star$ acts on
a polyform $\phi$ as
\eq{
\label{spinoraction}
\mathbb{X} \cdot \phi = \iota_X \phi + a \wedge \phi \, .
}
Because this action satisfies $\left(\mathbb{X}_1 \cdot \mathbb{X}_2 + \mathbb{X}_2 \cdot \mathbb{X}_1\right) \cdot \phi = 2 \, \cali(\mathbb{X},\mathbb{Y}) \phi $,
with the natural $(6,6)$-signature metric defined as
\eq{
\cali(\mathbb{X}_1,\mathbb{X}_2) = \frac{1}{2} \left( a_2(X_1) + a_1(X_2)\right) \, ,
}
it makes $T_M \oplus T_M^\star$ into a Clifford algebra and $\phi$ into an $SO(6,6)$-spinor.

Two compatible pure spinors $(\Psi_1,\Psi_2)$ define a generalized metric $(g,b)$ (with $g$ an ordinary metric and $b$ a 2-form)
as follows. First we define the null spaces $L_{1},L_{2} \subset T_M \oplus T^\star_M$ of $\Psi_1$ and $\Psi_2$ respectively,
i.e.\ $\mathbb{X} \in L_{1}$ if and only if $\mathbb{X} \cdot \Psi_1=0$ and analogously for $L_2$, and their
complex conjugates $\overline{L_{1}},\overline{L_{2}}$. Next we can define the spaces
\eq{
C_+ = (L_1 \cap L_2) \cup (\overline{L_1} \cap \overline{L_2}) \, , \qquad
C_- = (L_1 \cap \overline{L_2}) \cup (\overline{L_1} \cap L_2) \, .
}
It is possible to show that the elements of $C_+$ and $C_-$ have the form
\eq{
\label{genmetric}
\mathbb{X}_+ = (X,(g+b)X) \in C_+ \, , \qquad \mathbb{X}_- = (X,(-g+b)X) \in C_- \, ,
}
with $X \in T_M$. It is then easy to extract the sought for generalized metric $(g,b)$.
For the pure spinors defined from spinor bilinears as in \eqref{purespinors}, we find $b=0$.
In fact, in this picture the 2-form $b$ is completely absorbed in $H$ in \eqref{bulksusy}.
The vectors of $C_+$ act as $SO(6)$ gamma-matrices on the left, while those of $C_-$ act on the right.
By making a $b$-transform on these pure spinors, $\Psi_\pm \rightarrow e^b \Psi_\pm$ with $d b = H$,
it is possible to switch to an alternative picture where $H=0$ in \eqref{bulksusy}, while its information is completely contained
in the generalized metric $(g,b)$. We will not use the latter picture in this paper, because $b$ is generically
not globally defined, so that one needs to allow gauge transformations between local patches.

A number of useful properties of the spinor bilinears defined in \eqref{purespinors}
follow from the Fierz identity
\begin{equation}
M = \frac{1}{8} \sum_{l} \frac{1}{l!} {\rm Tr} \left(\gamma_{i_1 \ldots i_l} M \right)
\gamma^{i_l\ldots i_1} \, .
\end{equation}
Note the reversal of the indices and the appearance of the factor $8$ from tracing over the spinor
representation. Taking $M=\eta^{(1)}_+ \eta^{(2)\dagger}_\pm$ we find
that \eqref{purespinors} can be explicitly expanded as follows
\eq{
\Psi_{+i_1 \ldots i_k} = \frac{1}{8} \eta_+^{(2)\dagger} {\gamma}_{i_k \ldots i_1} \eta_+^{(1)} \, , \qquad
\Psi_{-i_1 \ldots i_k} = \frac{1}{8} \eta_-^{(2)\dagger} {\gamma}_{i_k \ldots i_1} \eta_+^{(1)} \, .
}
Taking instead $M$ to be
\begin{subequations}
\begin{align}
\label{3wedge}
V^{jkl} & = \gamma^{jkl} \slashchar{\Psi_\pm} \pm \slashchar{\Psi_\pm} \gamma^{jkl}
+ 3 \, \gamma^{[j} \slashchar{\Psi_\pm} \gamma^{kl]}
\pm 3 \, \gamma^{[kl} \slashchar{\Psi_\pm} \gamma^{j]} = 8 \, \slashchar{dy^j \wedge dy^k \wedge dy^l \wedge \Psi_\pm} \, , \\
V^{jk} & = \gamma^{jk} \slashchar{\Psi_\pm} + \slashchar{\Psi_\pm} \gamma^{jk}
\pm 2\, \gamma^{[j} \slashchar{\Psi_\pm} \gamma^{k]} = 4 \, \slashchar{dy^j \wedge dy^k \wedge \Psi_\pm} \, ,
\end{align}
\end{subequations}
respectively, we find
\begin{subequations}
\label{inversepurespinors}
\begin{align}
& a^2 \Omega^{(1)} |b|^2= -64 i \left.\Psi_- \wedge \alpha(\Psi_+)\right|_3 \, , \\
& |a|^4 \star \omega^{(1)} = -16 \left.\left(\Psi_+ \wedge \alpha(\bar{\Psi}_+) + \Psi_- \wedge \alpha(\bar{\Psi}_-)\right)\right|_{4} \, .
\end{align}
\end{subequations}
As a companion to \eqref{3wedge} one can also define
\eq{
W^{jkl} = \gamma^{jkl} \slashchar{\Psi_\pm} \mp \slashchar{\Psi_\pm} \gamma^{jkl}
+ 3 \, \gamma^{[j} \slashchar{\Psi_\pm} \gamma^{kl]}
\mp 3 \, \gamma^{[kl} \slashchar{\Psi_\pm} \gamma^{j]} = 8 \, \slashchar{\iota^j \iota^k \iota^l \Psi_\pm} \, ,
}
and show the following -- which will be useful in demonstrating the equation of motion for $H$ --
\begin{subequations}
\begin{align}
\text{Tr} \, (\slashchar{F}V^{jkl}) & = -64 \star \left( \Psi_\pm \wedge \alpha(\star F) \right)^{jkl} \, , \\
\label{eom3i}
\text{Tr} \, (\slashchar{F}W^{jkl}) & = -64 \star \left( F \wedge \star \; \alpha(\Psi_\pm) \right)^{jkl} \, ,
\end{align}
\end{subequations}
for any (poly)form $F$.

Furthermore, $\Psi_\pm$ satisfy
\eq{
\langle \Psi_\pm, \bar{\Psi}_\pm \rangle = - \frac{i}{8} |a|^2 |b|^2 \text{vol}_6 \, ,
}
so that the generalized Calabi-Yau property
\eq{
\label{CYcond}
\langle \Psi_+, \bar{\Psi}_+ \rangle = \langle \Psi_-, \bar{\Psi}_- \rangle
}
is automatically obeyed
for $SO(6,6)$-spinors created as spinor bilinears. They are also automatically
pure and compatible.
In addition,
they satisfy the following duality properties
\begin{equation}
\Psi_\pm = -i \star \alpha(\Psi_\pm) = \pm i \alpha(\star \Psi_\pm) \, .
\end{equation}
Finally, using \eqref{Mcondition6d} it is straightforward to show the following reality properties
\begin{subequations}
\label{purespinorrealprop}
\begin{align}
\slashchar{\bar{\Psi}_+} & = \eta^{(1)}_- \eta^{(2)\dagger}_-  \, , &
\slashchar{\alpha(\bar{\Psi}_+)} & = \eta^{(2)}_+ \eta^{(1)\dagger}_+  \, , \hspace{2cm} \\
\slashchar{\bar{\Psi}_-} & = - \eta^{(1)}_- \eta^{(2)\dagger}_+  \, , &
\slashchar{\alpha(\bar{\Psi}_-)} & = \eta^{(2)}_- \eta^{(1)\dagger}_+ \, . \hspace{2cm}
\end{align}
\end{subequations}

\section{Type II supergravity}
\label{sugra}

The bosonic content of type II supergravity consists of a metric $g$, a dilaton $\Phi$, an NSNS three-form $H$ and RR-fields
$F_{(n)}$. In the democratic formalism of \cite{democratic}, with double the number
 of  RR-fields, $n$ runs over $0,2,4,6,8,10$ in IIA and over $1,3,5,7,9$ in type IIB. In this paper $n$ will always indicate
the dimension of the RR-fields; for example $(-1)^n$ stands for $+1$ in type IIA and $-1$ in type IIB.
After deriving the equations of motion from the action the redundant RR-fields can then be removed by hand
by means of the duality condition
\eq{
\label{Fduality}
F_{(n)} = (-1)^{\frac{(n-1)(n-2)}{2}} \star_{10} F_{(10-n)} \Rightarrow F = (-1)^{n-1} \alpha(\star_{10} F)=\star_{10} \, \alpha(F) \, .
}
As in the above equation we will often collectively denote the RR-fields with the polyform $F=\sum_n F_{(n)}$. We also have doubled
the RR-potentials, collectively denoted by $C=\sum_n C_{(n-1)}$. In addition they
satisfy $F=d_H C$.\footnote{In the type IIA case with non-zero Romans mass parameter $m$ the
potentials are in fact defined by $F = d_H C + m e^{-B}$. In particular $F_{(0)}=m$. Also the Chern-Simons term
in the D-brane action has to be
adjusted accordingly. One can check that this does not change the analysis of this paper.}
Taking the compactification ansatz \eqref{Fansatz} into account,
the duality relation translates into
\eq{
\label{Fduality6d}
\tilde{F}_{(n)} = (-1)^{\frac{(n-1)(n-2)}{2}} \star_6 \hat{F}_{(6-n)} \Rightarrow \tilde{F} = (-1)^{n-1} \alpha(\star_6 \hat{F})=\star_6 \, \alpha(\hat{F}) \, .
}

The fermionic content consists
of a doublet of gravitino's $\psi_{M}$ and a doublet of dilatino's $\lambda$.
The components of the doublet are of different chirality
in type IIA and of the same chirality in type IIB.

The supersymmetry variation of the gravitino and dilatino doublet are given by
\begin{subequations}
\label{susyvar}
\begin{align}
\delta \psi_M & = D_{M} \epsilon = \left(\nabla_M +
\frac{1}{4} \slashchar{H_M} \calp
+ \frac{e^\Phi}{16} \sum_n \slashchar{F_{(n)}} \Gamma_M \calp_n \right) \epsilon \, , \\
\delta \lambda & = \left( \slashchar{\partial} \Phi + \frac{1}{2} \slashchar{H} \calp + \frac{e^\Phi}{8} \sum_n (-1)^n (5-n) \slashchar{F_{(n)}} \calp_n \right) \epsilon \, ,
\end{align}
\end{subequations}
with
\begin{subequations}
\begin{align}
& \text{IIA}: \; \calp = \Gamma_{(10)} \, , & & \text{IIB}: \; \calp = \sigma_3,\\
& \text{IIA}: \; \calp_n = - (-\Gamma_{(10)})^{\frac{n}{2}}\, , & & \text{IIB}: \; \calp_n = \sigma_1 \; \left(\frac{n+1}{2}\,\text{even}\right), \; i \sigma_2 \; \left(\frac{n+1}{2}\,\text{odd}\right) \, .
\end{align}
\end{subequations}

The Einstein equation (in the string frame),
its trace, and the dilaton equation of motion,
including the contribution from the `Dirac-Born-Infeld' part of the calibrated sources, D-branes and orientifolds,
read
\begin{subequations}
\begin{align}
\label{einstein}
& R_{MN} + g_{MN} \left( \frac{1}{8} H^2 + \frac{e^{2 \Phi}}{32} \sum_n (n-1)F_{(n)}^2 +\frac{1}{4} \left(\nabla^2\Phi-2(\partial\Phi)^2\right) \right)
\nonumber \\ 
& + 2 \nabla_M \partial_N \Phi
 - \frac{1}{2} H_{M} \cdot H_{N} -\frac{e^{2\Phi}}{4} \sum_n F_{(n)M} \cdot F_{(n)N} \nonumber \\
& - 2 \kappa_{10}^2 e^{2\Phi} \star \! \langle \sum_n \left(-\frac{1}{16} n g_{MN} + \frac{1}{2} g_{P(M} dx^P \otimes \iota_{N)}\right)\Psi_n,j_{\text{total}}\rangle
= 0 \, , \\
\label{einsteintrace}
& R -5(\partial \Phi)^2 +\frac{9}{2}\nabla^2\Phi- \frac{1}{4} H^2 - \frac{e^{2 \Phi}}{16} \sum_n (5-n) F_{(n)}^2
+\frac{\kappa_{10}^2 e^{2\Phi}}{4} \star \! \langle \sum_n n \Psi_n,j_{\text{total}}\rangle = 0 \, , \\
\label{dilaton}
& 2 R - H^2 + 8 (\nabla^2 \Phi - (\partial \Phi)^2) +(2 \kappa_{10}^2)e^{2\Phi} \star \! \langle \Psi,j_{\text{total}}\rangle = 0 \, ,
\end{align}
\end{subequations}
with $j_{\text{total}} = \sum_{\text{D}p} T_p j_{(\Sigma_p,\calf)} + \sum_{\text{O}p} T_{O_p} j_{(\Sigma_p)}$.
Finally, the Bianchi identities and equations of motion
for the RR-fields, including  the contribution from the `Chern-Simons' terms of the sources,
take the form
\begin{subequations}
\label{eomB}
\begin{align}
d_{-H} \star F & = 2 \kappa_{10}^2 \, \alpha(j_{\text{total}}) \, , \\
d_{H} F & = - 2 \kappa_{10}^2 \,  j_{\text{total}} \, ,
\end{align}
\end{subequations}
and the equation of motion for $H$
\eq{
\label{eomH}
d (e^{-2\Phi} \star \! H) - \frac{1}{2} \sum_n \star F_{(n)} \wedge F_{(n-2)} + \left. (2 \kappa_{10}^2) \Psi \wedge \alpha(j_{\text{total}})\right|_8=0 \, .
}
Note that the last term in the above equation comes from the Dirac-Born-Infeld term of the sources, while a careful analysis reveals that the
Chern-Simons contribution cancels with a contribution from the RR part of the bulk action upon using \eqref{eomB}.

\section{Conditions for $\caln=1$ compactifications to $M^{1,3}$}
\label{allconditions}

We collect here the complete conditions for an $\caln=1$
four-dimensional Minkowski background. The latter is described
by two complex polyforms $\Psi_1,\Psi_2$,
whereas  the (smeared or localized) sources are given by a real polyform $j_{\text{total}}$. Considered as spinors of the 12-dimensional
space $T_M \oplus T^\star_M$, $\Psi_1$ and $\Psi_2$ must be pure
i.e.\ their annihilator space must be maximal
(six-dimensional in the present case).

Every $SO(6,6)$ spinor and thus also $j_{\text{total}}$ can be written as a sum of pure spinors.
For each term the purity means that, roughly-speaking, it can be written as $\theta_p \wedge e^{-\calf}$
with the $p$-form $\theta_p$ decomposable in one-forms -- so that it can be interpreted as a single
D-brane or orientifold source. {}From a microscopic point of view one should require supersymmetry
for each of the sources, and thus the calibration constraints \eqref{calcond} for each individual
term.

The pure spinors $\Psi_1$ and $\Psi_2$ must satisfy
\begin{subequations}
\begin{align}
& \langle \Psi_1,\bar{\Psi}_1 \rangle = \langle \Psi_2,\bar{\Psi}_2 \rangle \neq 0 \, , \\
& \langle \Psi_1, \mathbb{X} \cdot \Psi_2 \rangle = \langle \bar{\Psi}_1, \mathbb{X} \cdot \Psi_2 \rangle = 0 \, , \quad \forall \mathbb{X} \in \Gamma(T_M \oplus T^\star_M) \, , \\
& g({\Psi_+,\Psi_-}) \; \text{positive-definite} \, .
\end{align}
\end{subequations}
See the discussion around \eqref{genmetric} for the prescription for finding the metric $g({\Psi_+,\Psi_-})$
associated with both pure spinors. Note that one has to explicitly check
the positive-definiteness of this metric.
It follows from these conditions that the two (almost) generalized complex structures associated to
$\Psi_+$ and $\Psi_-$ are commuting and the structure is $SU(3) \times SU(3)$.
These conditions, as well as
the condition of purity, are automatically satisfied in the case where $\Psi_+$
and $\Psi_-$ are constructed from spinor bilinears as in \eqref{purespinors}.

The calibration conditions for each D-brane and orientifold plane (for the latter $\calf=0$) read
\begin{subequations}
\label{calcond}
\begin{align}
\label{calcond1}
& \langle \Re \Psi_1, j_{(\Sigma,\calf)} \rangle = 0 \, , \\
\label{calcond2}
& \langle \Psi_2, \mathbb{X} \cdot j_{(\Sigma,\calf)} \rangle = 0 \, , \quad \forall \mathbb{X} \in \Gamma(T_M \oplus T^\star_M) \, , \\
\label{calcond3}
& \langle \Im \Psi_1, j_{(\Sigma,\calf)} \rangle/\text{vol}_6 > 0 \, .
\end{align}
\end{subequations}
As shown in \cite{gencal,lucasmyth}, for a localized D-brane source,
the calibration conditions are equivalent to the statement
that the source preserves the background supersymmetry \eqref{branesusy}. In addition,
they imply the equations of motion for the D-brane world-volume fields, provided one takes \eqref{diffsusyeq} into account.
In fact, supersymmetry will lead to the same equations also for smeared sources,
and then our argument that the supersymmetry conditions together with the form equations imply the
source-corrected Einstein and dilation equations holds regardless of whether the source is localized or smeared.

In addition, the differential supersymmetry conditions read:
\begin{subequations}
\label{diffsusyeq}
\begin{align}
d_H \left(e^{3A-\Phi} \Im \Psi_1 \right) & = \frac{e^{4A}}{16} \tilde{F} \, , \\
d_H \left(e^{A-\Phi} \Re \Psi_1 \right) & = 0 \, , \\
d_H \left(e^{2A-\Phi} \Psi_2 \right) & = 0 \, .
\end{align}
\end{subequations}
Finally we have the Bianchi identities for the form-fields:
\begin{subequations}
\label{bianchis}
\begin{align}
d_H \hat{F} & = - 2 \kappa_{10}^2 j_{\text{total}} \, , \\
d H & = 0 \, .
\end{align}
\end{subequations}
As shown in \cite{granaN1},
conditions \eqref{diffsusyeq} guarantee that the supersymmetry variation of
the gravitino and dilatino \eqref{susyvar} vanish,
so the background is supersymmetric. They also imply the equations of motion
for $\hat{F}$ and $H$ following section \ref{heom}.
Moreover, as shown in section \ref{integrability}, conditions \eqref{diffsusyeq}
together with the Bianchi identities \eqref{bianchis} imply
that the Einstein equation and the dilaton equation of motion are satisfied --
even in the presence of sources --  provided that the sources
satisfy the calibration conditions
\eqref{calcond}.

As shown in \cite{granascan},
equation \eqref{calcond3} together with the Bianchi identities
leads directly to the no-go theorem.
Indeed, suppose that all $T_p>0$ and non-zero fluxes then
\begin{multline}
\label{nogo}
0 \le \int_M \langle e^{3A-\Phi} \Im \Psi_1, \sum_{\text{sources}} T_p \, j_{(\Sigma,\calf)} \rangle
= -\frac{1}{4 \kappa_{10}^2} \int_M \langle e^{3A-\Phi} \Im \Psi_1, d_H \hat{F} \rangle \\
= -\frac{1}{4 \kappa_{10}^2} \int_M \langle d_H\left(e^{3A-\Phi} \Im \Psi_1 \right), \hat{F} \rangle
= -\frac{e^{4A}}{64 \kappa_{10}^2} \int_M \langle \tilde{F}, \hat{F} \rangle < 0 \, .
\end{multline}
It follows that at least one $T_p <0$, so we must have at least one orientifold.

\section{Nilmanifolds}
\label{nilmanifolds}

A nilmanifold has a basis of globally defined one-forms $e^a_L$, called left-invariant
one-forms, satisfying the Maurer-Cartan
relation
\eq{
\label{maurercartan}
d {e}^a_L = \frac{1}{2} f^a{}_{bc} \, {e}^b_L \wedge {e}^c_L \, ,
}
where $f^a{}_{bc}$ are the structure constants of the underlying nilpotent Lie-algebra.
The one-forms ${e}^a_L$ are not necessarily a vielbein, although a simple
choice for the vielbein $e^a$ (and corresponding metric) could indeed be (a warping of) $e^a_L$.
One can always make a nilmanifold compact
by dividing by a discrete group $\Gamma$. Moreover, when restricting to left-invariant structures,
i.e.\ structures with constant coefficients in the basis
of the left-invariant one-forms, the analysis does not depend on the choice of $\Gamma$.
For a nilpotent algebra, there is always a choice of ${e}^a_L$s
such that $f^a{}_{bc}$ is integer and non-zero only if $b<a,c<a$. With such a choice,
notation such as $(0,0,0,0,13+42,14+23)$ stands for a nilmanifold with $d{e}^5_L= {e}^1_L \wedge {e}^3_L +  {e}^4_L \wedge {e}^2_L$,
$d{e}^6_L= {e}^1_L \wedge {e}^4_L +  {e}^2_L \wedge {e}^3_L$, and all other $de^a_L$ zero (this is the Iwasawa manifold).
There are 34 isomorphism classes of six-dimensional, simply-connected, nilpotent Lie-groups, for which in
the present paper we use the
 numbering of table 4 of \cite{granascan}.
In the physics literature (compactified) nilmanifolds are also
called twisted tori, because they can be regarded as iterated torus bundles.


\listoftables       
\listoffigures      

\end{document}